\documentclass[aps,pra,twocolumn,superscriptaddress,amsmath,amssymb,10pt]{revtex4-2}

\usepackage{graphicx}
\usepackage{dcolumn}
\usepackage{bm}
\usepackage{physics}
\usepackage{amsfonts, amsmath}
\usepackage{dsfont}
\usepackage{amsthm}
\usepackage{hyperref}
\usepackage{bm}
\usepackage{bbold}
\usepackage{color}
\usepackage{xcolor}
\usepackage{lipsum,babel}
\usepackage[normalem]{ulem}

\DeclareMathOperator*{\V}{\mathbb{V}}

\newcommand{\cor}[1]{{\color{red}{#1}}}

\begin{document}

\definecolor{navy}{RGB}{46,72,102}
\definecolor{pink}{RGB}{219,48,122}
\definecolor{grey}{RGB}{184,184,184}
\definecolor{yellow}{RGB}{255,192,0}
\definecolor{grey1}{RGB}{217,217,217}
\definecolor{grey2}{RGB}{166,166,166}
\definecolor{grey3}{RGB}{89,89,89}
\definecolor{red}{RGB}{255,0,0}

\preprint{APS/123-QED}

\title{Virtual purification complements quantum error correction in quantum metrology}

\author{Hyukgun Kwon}
\email{kwon37hg@sejong.ac.kr}
\affiliation{Department of Physics and Astronomy, Sejong University, 209 Neungdong-ro Gwangjin-gu, Seoul 05006, Republic of Korea}
\affiliation{Pritzker School of Molecular Engineering, University of Chicago, Chicago, Illinois 60637, USA}

\author{Changhun Oh}
\affiliation{Department of Physics, Korea Advanced Institute of Science and Technology, Daejeon 34141, Republic of Korea}

\author{Youngrong Lim}
\affiliation{Department of Physics, Chungbuk National University, Cheongju, Chungbuk, 28644, Republic of Korea}
\affiliation{School of Computational Sciences, Korea Institute for Advanced Study, Seoul 02455, Republic of Korea}

\author{Hyunseok Jeong}
\affiliation{Department of Physics and Astronomy, Seoul National University, Seoul 08826, Republic of Korea}

\author{Seung-Woo Lee}
\email{swleego@gmail.com}
\affiliation{Department of Physics, Pohang University of Science and Technology (POSTECH), Pohang 37673, Republic of Korea}

\author{Liang Jiang}
\email{liangjiang@uchicago.edu}
\affiliation{Pritzker School of Molecular Engineering, University of Chicago, Chicago, Illinois 60637, USA}

\begin{abstract}
Quantum resources enable one to achieve quantum-enhanced estimation sensitivity beyond its classical counterpart. Many studies mainly focus on reducing statistical error, under the assumption that one can always set an unbiased estimator. However, setting an unbiased estimator is not always feasible, especially when one cannot fully characterize noise. Such incomplete noise characterization induces a bias and eventually makes it impossible to attain the enhanced-estimation. In this work, we explore two systematic approaches; quantum error correction (QEC) and the virtual purification (VP) to reduce the bias, and compare their performance. First, we show that when the noise is indistinguishable from the signal, QEC cannot reduce the bias since it is impossible to construct a QEC code that corrects the noise while preserving the signal. We then show that VP can mitigate indistinguishable error that eventually enable a more accurate estimation compared to QEC. Our findings reveal that VP offers a robust alternative to QEC in scenarios where indistinguishable errors pose significant challenges. We then demonstrate that VP with a stabilizer state probe can efficiently suppress the bias under local depolarizing noise, thereby yielding a significant improvement in estimation performance compared to the QEC-based approach.
\end{abstract}
              
\maketitle

\section{Introduction}
Quantum metrology enhances the sensitivity of signal estimation by exploiting entangled resources to reach the Heisenberg scaling beyond the standard quantum limit \cite{caves1981quantum,     baumgratz2016quantum, stbs-giovannetti2006quantum, Giovannetti2011, braun2018quantum, Pirandola2018}. This enables significant improvements in applications such as magnetometry, interferometry, gravitational wave detection, and atomic clocks, providing a potential for advancing various sensing technologies \cite{PhysRevLett.71.1355,taylor2008high, giovannetti2001quantum}.
However, these advantages are often diminished due to the noise and decoherence from environmental interactions \cite{noi-demkowicz2009quantum, noi-demkowicz2012elusive, noi-escher2011general, noi-huelga1997improvement, noi-yamamoto2021error}, limiting the precision and accuracy of estimation, which restricts its practical applications.

To tackle these challenges, while various schemes have been developed \cite{technique-albarelli2018restoring, technique-johnsson2020geometric, technique-liu2017quantum, technique-liu2023optimal, technique-maze2008nanoscale, technique-viola1999dynamical}, quantum error correction (QEC)--a standard way of dealing with noise mostly in quantum computing--has been considered for quantum metrology \cite{qec-PhysRevLett.112.080801, qec-PhysRevLett.112.150801, qec-PhysRevLett.112.150802, qec-PhysRevLett.122.040502, qec-PhysRevX.7.041009, qec-zhou2018achieving, qec-PRXQuantum.2.010343,  qec-zhuang2020distributed, qec-kwon2025restoring}. Early studies have shown that QEC can restore the Heisenberg scaling when the noise is perpendicular to the signal \cite{qec-PhysRevLett.112.080801, qec-PhysRevLett.112.150801, qec-PhysRevLett.112.150802}. A general criterion known as the \textit{Hamiltonian-not-in-Kraus (or Lindblad)-span} (HNKS) condition was then established to assess the feasibility of achieving the Heisenberg scaling with QEC \cite{qec-PhysRevX.7.041009, qec-zhou2018achieving, qec-PRXQuantum.2.010343}. When the HNKS condition is satisfied, noise are distinguishable from signal. This distinguishability enables the construction of a QEC code that selectively corrects noise while preserving the signal. As a result, Heisenberg scaling can be achieved by the application of QEC \cite{qec-PhysRevX.7.041009, qec-zhou2018achieving, qec-PhysRevLett.122.040502}. In contrast, when the HNKS condition is violated, the noise becomes indistinguishable from the signal, making it impossible to construct a QEC code that corrects the noise without simultaneously disturbing the signal. Consequently, the recovery of enhanced estimation is precluded, even in an idealized noiseless QEC setting supplemented with ancillary systems.

Beyond this fundamental limitation, much of the existing literatures have focused on precision limits in the presence of noise or on QEC-assisted metrology under the assumption that the noise can be accurately characterized. This assumption enables the use of asymptotically unbiased estimators, such as the maximum likelihood estimator (MLE) \cite{crb-fisher1925theory, crb-cramer1999mathematical, crb-lehmann2006theory, crb-braunstein1992large}. However, accurate noise characterization may not always be feasible in practice, thereby posing significant challenges to these approaches.
As a consequence, recent works have begun to focus on biases induced by unknown noise \cite{noi-yamamoto2021error, qem-kwon2024efficacy, bias-kwon2025criteria}. When the noise exhibits unknown characteristics (hereafter referred to as \textit{unknown noise}), such as uncertain noise parameters or, more critically, an unknown noise type, the construction of an unbiased estimator may no longer be feasible, inevitably inducing a bias in the estimator. Notably, as the sample size grows, the bias tends to outweigh the statistical error and becomes the predominant source of the total estimation error. This bias dominance ultimately prevents the attainment of Heisenberg scaling \cite{noi-yamamoto2021error, qem-kwon2024efficacy, biasqec-rojkov2022bias, bias-kwon2025criteria, supple}. In this regime, the bias sets a fundamental lower bound on the achievable estimation error; consequently, reducing the bias becomes the primary task for enhancing estimation performance.

Meanwhile, virtual purification (VP), a quantum error mitigation technique---originally developed for improving expectation value estimation of Pauli operator \cite{vp-PhysRevX.11.031057, vp-PhysRevX.11.041036} in the presence of incoherent noise---has been applied to reduce the effect of noise in quantum metrology. Specifically, it was recently shown that VP can suppress the bias caused by unknown noise due to that VP works without necessitating any information about noise, enabling more accurate estimation in noisy metrology \cite{noi-yamamoto2021error, qem-kwon2024efficacy}. Moreover, VP efficiently reduces bias even if noise fluctuates between measurements \cite{vp-PhysRevX.11.031057, vp-PhysRevX.11.041036,noi-yamamoto2021error}. However, VP has been applied only in regimes where the signal is sufficiently small, allowing signal estimation to be mapped onto expectation value estimation \cite{noi-yamamoto2021error, qem-kwon2024efficacy}. 

Despite the recent advancements of both QEC and VP, challenges still remain to fully handle the effects of noise in practical scenarios of quantum metrology. In addition, no comparative analysis between QEC and VP has been explored yet from a quantum metrology perspective. In this work, we fill this gap by evaluating and comparing their performance in realistic noisy quantum metrology. 
We consider the setting of canonical phase estimation---representative applications to magnetometry, interferometry, and quantum clocks---using stabilizer states as quantum probes, Pauli measurements, and MLE-based estimation in the presence of Pauli noise.
We first show how a bias occurs when the noise is not fully characterized and one disregards it. We then show that even an ideal QEC setup fails to correct local Pauli Z errors that are indistinguishable from the signal and induce bias, while VP can mitigate indistinguishable errors from the signal resulting in more accurate estimations than QEC in the same estimation scenario. In contrast to QEC, VP can effectively purify the noisy state, resulting in the amplification of its dominant eigenvector (which coincides with the noise-free state in our settings) and suppression of the error components induced by Pauli noise, irrespective of the distinguishability between noise and signal.

It is worth noting that the application of VP entails an inherent trade-off between bias reduction and statistical error. While increasing the number of copies can in principle further suppress the bias, it has been discussed that this may come at the price of increased statistical error \cite{qem-takagi2022fundamental, qem-PhysRevLett.131.210601, qem-PhysRevLett.131.210602, vp-PhysRevX.11.031057, vp-PhysRevX.11.041036}. At the same time, it has been shown that substantial error mitigation can already be achieved using VP with a small number of copies \cite{vp-PhysRevX.11.041036}. In this work, we thus focus on VP involving only two copies of the noisy state, which is sufficient to achieve a significant reduction of the bias while keeping the statistical error modest. We emphasize that, in the large-sample regime, the bias tends to become the dominant contribution to the total estimation error as the statistical error scales inversely with the sample size. Consequently, the reduction in bias compensates for the modest increase in statistical error, resulting in an overall reduction of the total estimation error and rendering VP particularly advantageous for quantum metrology.

Finally, we demonstrate that VP with a quantum probe in a stabilizer state can efficiently mitigate the bias induced by local Pauli noise in quantum metrology. Numerical simulations show substantial bias suppression when employing $5$-qubit GHZ state or the logical zero state of $7$-qubit Steane code as the probe state, resulting in a significant enhancement of the total estimation performance in the large sample limit. This implies that applying VP along with QEC-encoded probe states is effective against the unknown noise that QEC solely cannot address. Moreover, this approach is resource-efficient compared to QEC as it does not require repeated QEC cycles and ancillary systems. Our findings highlight that VP can complement QEC in noisy quantum metrology where error distinguishability poses significant challenges.

\begin{figure}[t]
    \centering
    \includegraphics[width=0.95\linewidth]{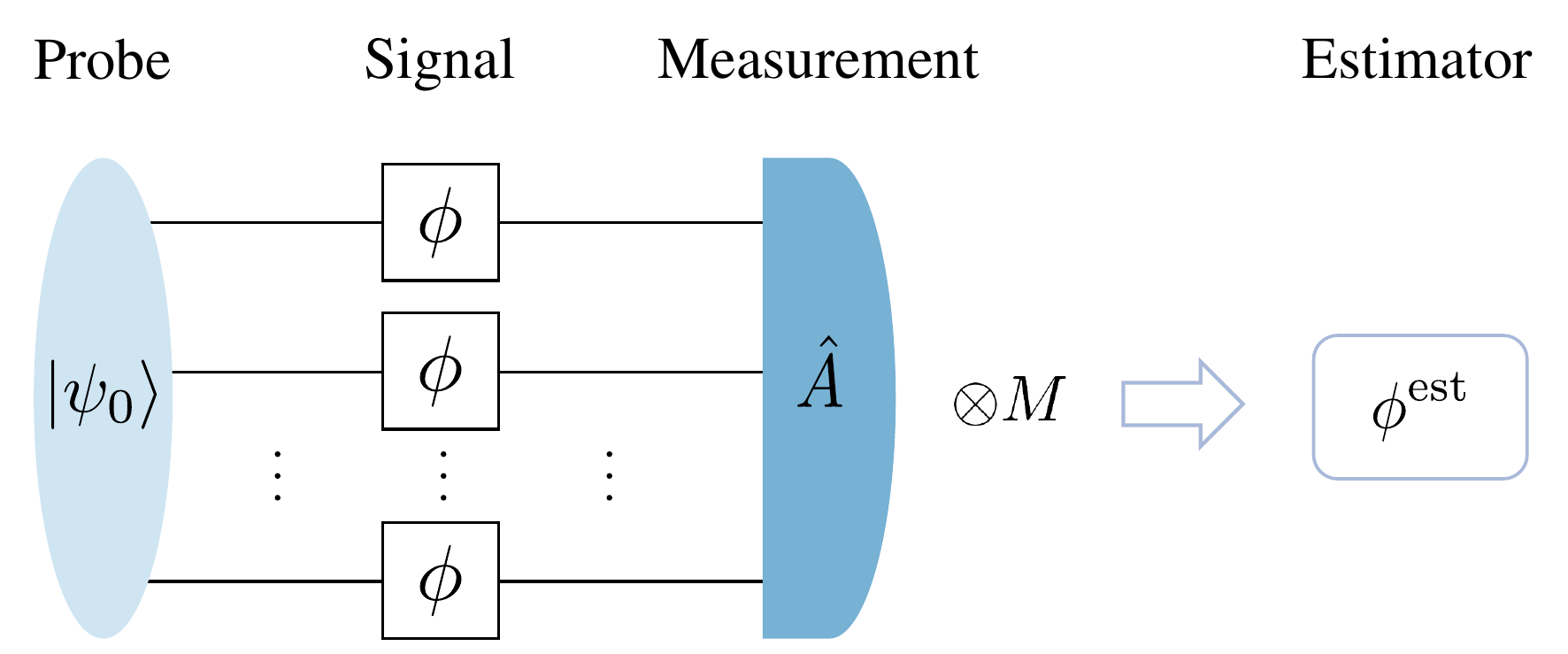}
    \caption{To estimate the signal $\phi$, an $N$-qubit quantum probe $\ket{\psi_{0}}$ is prepared and the probe imprints an identical signal $\phi$ into each qubit mode. The corresponding signal state is then measured, and this process is repeated $M$ times to gather $M$ measurement outcomes. An estimator $\phi^{\mathrm{est}}$ is then derived from the outcomes, providing an estimated value of $\phi$.} 
    \label{fig:metrology}
\end{figure}

\section{Canonical phase estimation in the presence of unknown noise}
\subsection{Canonical phase estimation}
Let us begin by considering the canonical phase estimation as illustrated in Fig.~\ref{fig:metrology}, where an unknown signal $\phi$ is imprinted by the unitary operation
\begin{align}
    \hat{U}(\phi)=\exp\left(-i\frac{\phi}{2}\sum_{i=1}^{N}\hat{Z}^{(i)}\right)=\exp\left(-i\frac{\phi}{2}\hat{H}\right).
\end{align}
Here, $\hat{Z}^{(i)}$ is the Pauli $Z$ operator acting on $i$th qubit and $\hat{H}:= \sum_{i=1}^{N}\hat{Z}^{(i)}$ is the signal Hamiltonian. We refer to the Pauli operator that acts on a single qubit as a \textit{single Pauli operator}. To estimate $\phi$, a quantum probe $\ket{\psi_{0}}$ is prepared and evolves under the unitary dynamics, resulting in the signal state $\ket{\psi(\phi)}:=\hat{U}(\phi)\ket{\psi_{0}}$. The signal state is then measured by the eigenbasis of an observable $\hat{A}$. Repeating this process $M$ times yields a series of measurement outcomes, denoted as $\vec{A}=(A_{1},A_{2},\cdots, A_{M})$. We note that an estimation performance depends on the choice of $\hat{A}$. However, the optimal measurement generally depends on the signal and can involve a highly entangled measurement, which is challenging to realize. Here, we thus focus on more practical scenarios by employing Pauli measurements for estimation. Given the measurement outcomes, we construct an estimator $\phi^{\mathrm{est}}=\phi^{\mathrm{est}}(\vec{A})$. The estimation performance is then characterized by the mean squared error defined as
\begin{align}
    \left<(\phi^{\mathrm{est}}-\phi)^{2}\right>=\left(\left<\phi^{\mathrm{est}}\right>-\phi\right)^{2}+\left<(\phi^{\mathrm{est}})^{2}\right>-\left<\phi^{\mathrm{est}}\right>^{2}, \label{MSE}
\end{align}
where $\langle \cdot \rangle$ denotes the average over all possible measurement outcomes. The first term $\left<\phi^{\mathrm{est}}\right>-\phi$ represents the \textit{bias}, and the second term is the \textit{statistical error} that characterizes the estimation accuracy and precision respectively. Notably, according to the central limit theorem, the statistical error, which is the variance of an estimator, scales inversely with the number of samples $M$. 

We adopt MLE, a widely used estimator in quantum metrology, due to its unbiased nature $\left<\phi^{\mathrm{est}}\right>=\phi$ and its ability to asymptotically achieve the minimum statistical error in the large sample limit $M \to \infty$ \cite{crb-fisher1925theory, crb-cramer1999mathematical, crb-lehmann2006theory, crb-braunstein1992large}. 
The MLE is defined as the parameter value that maximizes the probability of observing the outcomes $\vec{A}$ \cite{crb-fisher1925theory, crb-cramer1999mathematical, crb-lehmann2006theory, crb-braunstein1992large}. 
In the canonical phase estimation protocol based on Pauli measurements, the MLE admits a particular simple closed-form expression (see Methods Sec. A), 
\begin{align}
    \phi^{\mathrm{est}}=\mu^{-1}(\bar{A}), \label{mle}
\end{align}
where $\bar{A}$ denotes empirical mean of the outcomes defined as 
\begin{align}
    \bar{A} := \frac{1}{M}\sum_{i=1}^{M} A_{i},
\end{align}
and
\begin{align}
    \mu(\phi):=\Tr\left[\hat{A}\dyad{\psi(\phi)}\right]=\left<\bar{A}\right> 
\end{align}
is the expectation value of the observable $\hat{A}$ with respect to the signal state $\ket{\psi(\phi)}$. The function $\mu^{-1}$ denotes the inverse of $\mu(\phi)$, assumed to be well defined in the relevant parameter regime. (See Methods Sec. A. and SM. Sec. S1 for the details.)
We emphasize that Eq. \eqref{mle} formalizes an intuitive estimation rule: the unknown parameter $\phi$ is inferred by equating the experimentally observed sample mean $\bar{A}$ to the theoretical expectation value $\mu(\phi)$. Indeed, for repeated Pauli measurements, the law of large numbers guarantees that the empirical average $\bar{A}$ converges to $\mu(\phi)$ as the number of measurement repetitions $M$ increases. Consequently, the MLE becomes an unbiased estimator in the asymptotic limit $M\to\infty$:
\begin{align}
    \left<\phi^{\mathrm{est}}\right>=\left<\mu^{-1}(\bar{A})\right>=\phi. 
\end{align}
In the asymptotic limit, the estimator is asymptotically unbiased, and the mean squared error is entirely determined by its statistical error
\begin{align}
\left<(\phi^{\mathrm{est}}-\phi)^{2}\right>=\frac{1}{M}\left(\frac{1-\mu(\phi)^{2}}{\pdv{\mu(\phi)}{\phi}}\right)^{2}.\label{idealstat}
\end{align}
Consequently, minimizing the statistical error of the measurement outcomes directly enhances the estimation performance, which is the primary focus of most studies in quantum metrology. (A detailed derivation of Eq. \eqref{idealstat} is provided in the Methods Sec. A.)

\subsection{Bias induced from unknown noise}
\begin{figure}[t]
    \centering
    \includegraphics[width=0.99\linewidth]{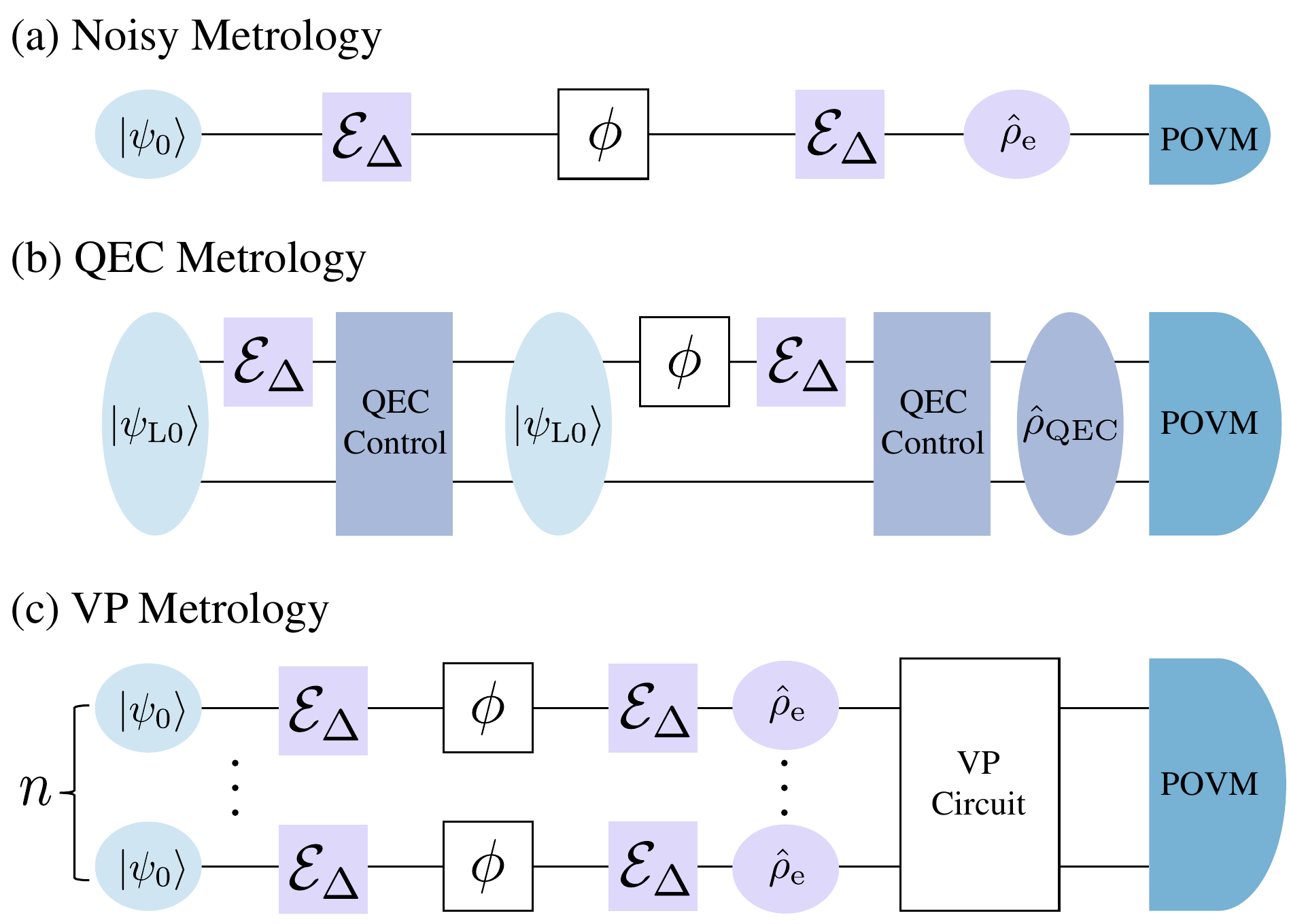}
    \caption{(a)-(c) Schematic illustrations of noisy quantum metrology, QEC-assisted metrology, and VP-assisted metrology are shown, respectively. We consider an $N$-qubit quantum probe prepared in an initial state
$\ket{\psi_{0}}$, where the same unknown signal phase is imprinted
identically onto each qubit; this signal imprinting process is represented
by the boxed phase shift.
(a) In the noisy quantum metrology scheme, Pauli noise acts on the probe
both before and after the signal imprinting process.
(b) In the quantum error correction (QEC)-assisted scheme, ideal QEC control
is applied immediately after the occurrence of Pauli noise.
To illustrate that QEC cannot efficiently correct the bias induced by
unknown Pauli noise, we assume that the QEC procedure perfectly corrects
the effect of the first noise process.
(c) In the virtual purification (VP)-assisted scheme, multiple identical
copies of the noisy quantum state are prepared simultaneously, and VP is
applied to mitigate the bias arising from the noise.}    
     
    \label{fig:overall}
\end{figure}

In more realistic settings, noise is inevitably present during the estimation process. Let us first illustrate how bias arises when the same estimator
defined in Eq. \eqref{mle} is exploited in the presence of unknown noise. We then show that such a noisy signal estimation can be mapped to a noisy expectation value estimation. Consequently, VP, originally developed for the accurate expectation value estimation in noisy quantum computation, can be leveraged to achieve more accurate signal estimation.

We first consider the case in which noise modifies the ideal signal state, such that the ideal signal state $\ket{\psi(\phi)}$ is replaced by a noisy state, denoted by $\hat{\rho}_{\mathrm{e}}(\phi)$. As a result, repeated measurements yield an empirical sample mean $\bar{A}_{\mathrm{e}}$ obtained from $\hat{\rho}_{\mathrm{e}}(\phi)$, rather than the ideal mean $\bar{A}$. Although noise degrades the statistical error compared to the ideal case, it remains possible to construct the MLE in direct analogy with Eq. \eqref{mle}, when the noisy state $\hat{\rho}_{\mathrm{e}}$ can be fully characterized. However, when the noise cannot be fully characterized, or equivalently when the noisy state $\hat{\rho}_{\mathrm{e}}$ is not fully characterized, the theoretical reference function $\mu_{\mathrm{e}}(\phi):=\left<\bar{A}_{\mathrm{e}}\right>$ becomes unavailable. As a consequence, constructing MLE analogous to Eq. \eqref{mle} is, in general, not feasible. In particular, the absence of the theoretical reference function $\mu_{\mathrm{e}}(\phi)$ precludes direction inversion required in Eq. \eqref{mle}, leaving only the empirically accessible sample mean $\bar{A}_{\mathrm{e}}$. Instead, a natural alternative is to use the estimator defined in Eq. \eqref{mle}, i.e.,
\begin{align}
    \phi^{\mathrm{est}}_{\mathrm{e}}:= \mu^{-1}(\bar{A}_{\mathrm{e}}), \label{esterror}
\end{align}
which leverages the known theoretical reference function $\mu(\phi)$ and the given empirical sample mean $\bar{A}_{\mathrm{e}}$. In this case, however, a systematic bias is inevitably introduced due to the unknown form of $\mu_{\mathrm{e}}(\phi)$ even in the large sample limit, that is given by
\begin{align}
    B_{\mathrm{e}}:=\left<\phi^{\mathrm{est}}_{\mathrm{e}}\right>-\phi = \mu^{-1}(\mu_{\mathrm{e}}(\phi))-\mu^{-1}(\mu(\phi)). \label{biasgeneral}
\end{align}
(A detailed derivation and the explicit expression of Eq. \eqref{biasgeneral} are provided in the Methods Sec. B.)
Equation~\eqref{biasgeneral} shows that the bias originates from the discrepancy between the noisy and ideal expectation values $(\mu_{\mathrm{e}} - \mu)$. This implies that suppressing the discrepancy $(\mu_{\mathrm{e}} - \mu)$ directly reduces the bias $B_{\mathrm{e}}$, thereby enhancing the accuracy of the parameter estimation.

\subsection{Pauli noise}
Next, we consider the case in which independent and identically distributed Pauli noise (IIDP) is the dominant noise source during the estimation process, affecting both sides of the signal imprinting. (See. Fig. \ref{fig:overall} (a).) The resulting noisy state can be described as 
\begin{align}
    \hat{\rho}_{\mathrm{e}}(\phi,\Delta) := \mathcal{E}_{\Delta}\circ\mathcal{U}_{\phi}\circ\mathcal{E}_{\Delta}(\dyad{\psi_{0}}). 
\end{align}
where
\begin{align}
    \mathcal{U}_{\phi}(\hat{\rho})=\hat{U}(\phi)\hat{\rho}\hat{U}^{\dagger}(\phi),~~ \mathcal{E}_{\Delta}(\hat{\rho})=\mathcal{E}^{(1)}_{\Delta} \circ \mathcal{E}^{(2)}_{\Delta} \circ \cdots \circ \mathcal{E}^{(N)}_{\Delta}(\hat{\rho}). \label{IIDP}
\end{align}
Here $\mathcal{E}^{(i)}$ represents the Pauli channel acting on $i$th qubit, defined as
\begin{align}
    \mathcal{E}^{(i)}_{\Delta}(\hat{\rho})=p_{I}\hat{\rho}+p_{x}\hat{X}^{(i)}\hat{\rho}\hat{X}^{(i)}+p_{y}\hat{Y}^{(i)}\hat{\rho}\hat{Y}^{(i)}+p_{z}\hat{Z}^{(i)}\hat{\rho}\hat{Z}^{(i)}, \label{IIDPnoise}
\end{align}
where $\hat{X}^{(i)}$, $\hat{Y}^{(i)}$, and $\hat{Z}^{(i)}$ being the single Pauli $X$, $Y$, and $Z$ operators that acting on $i$th qubit, respectively, and the probabilities satisfying $p_{I} + p_{x} + p_{y} + p_{z} = 1$. We assume that the IIDP noise is characterized by its noise strength $\Delta$, with the probabilities $p_{I},p_{x},p_{y},p_{z}$ being functions of $\Delta$. In addition, we consider mild noise limit where $\Delta$ is small and the noise distribution that satisfies 
\begin{align}
    p_{I}=O(1),~p_{x},p_{y}=\Theta(\Delta),~p_{z}=O(\Delta),
\end{align}
or
\begin{align}
    p_{I}=O(1),~p_{x},p_{y}=O(\Delta),~  p_{z}=\Theta(\Delta).
\end{align}
This noise model captures both local depolarizing and dephasing errors. (The explicit form of $\hat{\rho}_{\mathrm{e}}$ can be found in the Supplemental Material (SM) Sec. S2.) 
In this setting, if the noise strength $\Delta$ cannot be characterized—or, more generally, if the noise is not even known to be Pauli channel—the resulting estimator becomes systematically biased. According to Eq.~\eqref{biasgeneral} and using the relation 
\begin{align}
    \mu_{\mathrm{e}}(\phi,\Delta)-\mu(\phi)=O(\Delta),
\end{align}
which follows from a Taylor expansion in $\Delta$ where $\mu_{\mathrm{e}}(\phi,\Delta):=\mathrm{Tr}\left[\hat{A}\hat{\rho}_{\mathrm{e}}(\phi,\Delta)\right]$, we find that the corresponding bias is \cite{bigO}
\begin{align}
    B_{\mathrm{e}}(\phi, \Delta)= O(\Delta).  \label{biase}
\end{align}
(A detailed derivation and the explicit expression of Eq. \eqref{biase} are provided in the Methods Sec. B.)

\begin{figure}[t]
    \centering
    \includegraphics[width=0.95\linewidth]{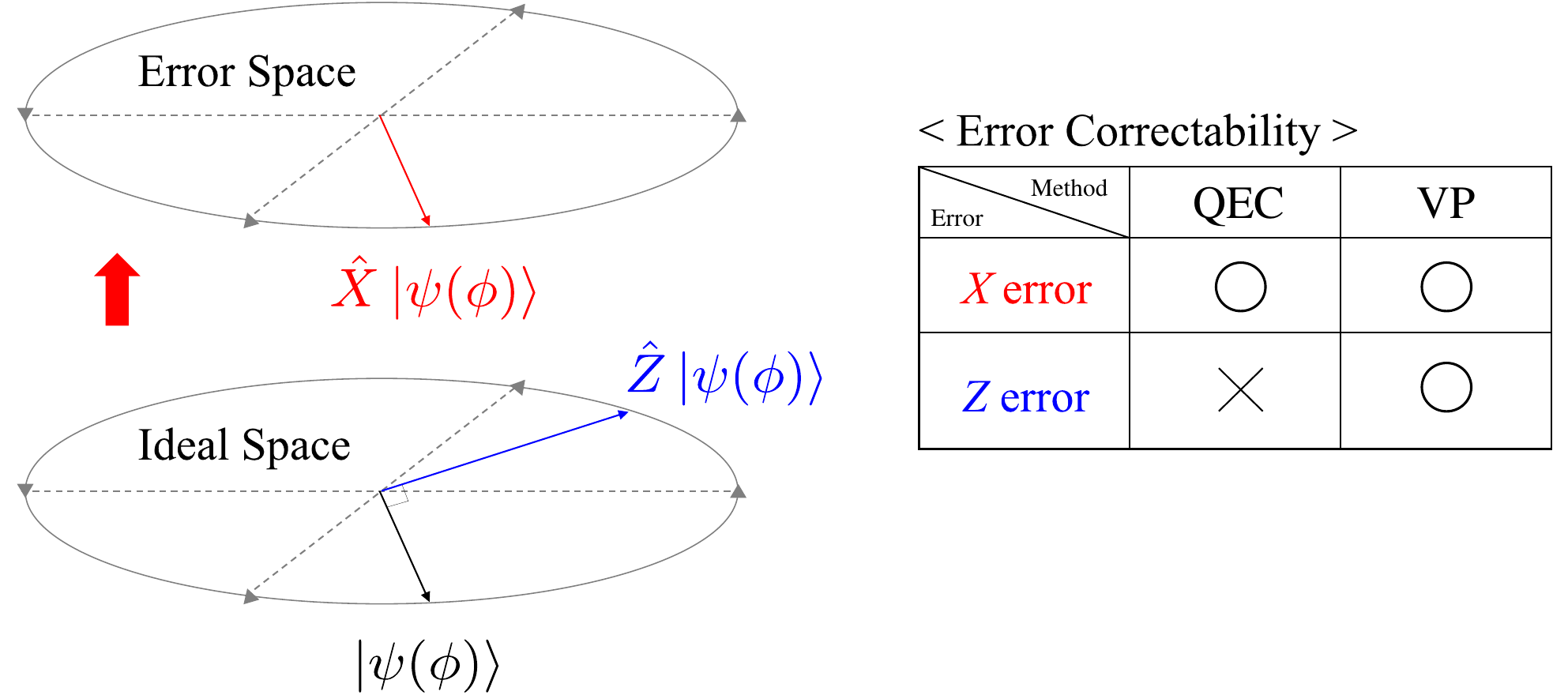}
    \caption{$\ket{\psi(\phi)}$ is the signal state and the ideal space is the space in which the signal state lies in. While the single Pauli $\hat{X}$ error can be suppressed by both methods, the single Pauli $\hat{Z}$ error can only be mitigated by virtual purification method.
} 
    \label{fig:errorclass}
\end{figure}

\section{Limitation of QEC}
Next, let us inspect whether QEC enables unbiased estimation. We show in what follows that even ideal QEC setup fails to restore the unbiased estimation since the Pauli $Z$ error is indistinguishable from the signal. As a result, no QEC code can simultaneously correct the error while preserving the signal.
We assume noiseless operations for encoding, syndrome measurements, and recovery are always possible, and noiseless ancillae qubits are available. However, the signal $\phi$ cannot be imprinted in the ancilla mode. Throughout the QEC section, we employ a shorthand notation $\hat{\Omega}$ to denote an operator of the form $\hat{\Omega} \otimes \hat{I}_{\mathrm{A}}$, where $\hat{\Omega}$ is an operator acting on $N$-qubit system and $\hat{I}_{\mathrm{A}}$ is the identity operator on the ancilla system. We denote the code space as $\mathcal{C}$ (in Fig. \ref{fig:errorclass}, we call the code space as ideal space), the logical quantum probe as $\ket{\psi_{\mathrm{L}0}} \in \mathcal{C}$, and the signal state as $\ket{\psi_{\mathrm{L}}(\phi)} := \hat{U}(\phi)\ket{\psi_{\mathrm{L}0}}$. Additionally, we assume that any noise present before the signal imprinting can be completely corrected by QEC, allowing the use of a noiseless quantum probe $\ket{\psi_{\mathrm{L}0}}$. A schematic of the QEC case is shown in Fig.~\ref{fig:overall} (b). Consequently, the error state after signal imprinting can be described as $\mathcal{E}_{\Delta}\left(\dyad{\psi_{\mathrm{L}}}\right)=\sum_{i}\hat{E}_{i}\dyad{\psi_{\mathrm{L}}}\hat{E}^{\dagger}_{i}$ where $\{\hat{E}\}_{i}$ is the Kraus operator set for the IIDP noise defined in Eq. \eqref{IIDP}. QEC is then applied to this error state.

To inspect the efficacy of QEC, we consider the QEC scheme that satisfies the following conditions introduced in Ref. \cite{qec-PhysRevLett.112.150802}: (C1) $[\hat{\Pi}_{\mathcal{C}},\hat{H}]=0$ where $\hat{\Pi}_{\mathcal{C}}$ is the projection operator on the code space. This ensures that the signal state $\ket{\psi_{\mathrm{L}}}$ is always in the code space when an error does not occur. (C2) $\hat{\Pi}_{\mathcal{C}}\hat{E}^{\dagger}_{i}\hat{E}_{j}\hat{\Pi}_{\mathcal{C}}=h_{ij}\hat{\Pi}_{\mathcal{C}}~\forall i,j$ where $h_{ij}$ is a hermitian matrix. This is the Knill-Laflamme condition, which implies the existence of a recovery map that restores the noiseless state from the noise. (C3) The maximum quantum Fisher information among all possible quantum states in the code space must be larger than $0$: $\max_{\ket{\Psi_{\mathrm{L}}} \in \mathcal{C}}\left[\langle \Psi_{\mathrm{L}} \vert \hat{H}^{2} \vert \Psi_{\mathrm{L}} \rangle-\langle \Psi_{\mathrm{L}} \vert \hat{H} \vert \Psi_{\mathrm{L}} \rangle^{2} \right]> 0$. 
If this condition is violated, any quantum state in the corresponding code space cannot imprint the signal, rendering signal estimation impossible.

It has been shown that when the noise is limited to single Pauli $X$ errors, one can construct a code that satisfies all conditions (C1)–(C3) \cite{qec-PhysRevLett.112.150802} and perfectly recovers the ideal state, thereby enabling unbiased parameter estimation. 
However, it can be shown that no (QEC) scheme can simultaneously correct all single Pauli $Z$ error while satisfying (C3). The intuitive explanation is as follows. If one constructs a QEC code that satisfies the Knill–Laflamme condition in (C2), the code can efficiently correct Pauli $Z$ errors. However, the signal Hamiltonian $\hat{H}$ also consists of a sum of single Pauli $Z$ operators. Consequently, a QEC procedure that corrects Pauli $Z$ errors also removes the signal imprinted by $\hat{H}$. As a result, no information about the parameter $\phi$ remains, and the quantum Fisher information is $0$. Equivalently, condition (C3) cannot be satisfied. For the same reason, no QEC scheme can simultaneously correct all single Pauli $X$ and $Y$ errors while satisfying (C3). (Detailed proofs of these statements are provided in the Methods Sec. D.)

As a consequence, under the IIDP noise with noise distribution that satisfies $p_{I}=O(1)$, $p_{x},p_{y}=\Theta(\Delta)$ and $p_{z}=O(\Delta)$ or $p_{I}=O(1)$, $p_{x},p_{y}=O(\Delta)$ and $p_{z}=\Theta(\Delta)$, the error-corrected state is also a noisy state that is different from ideal state. Therefore, in analogous to Eq. \eqref{biase}, the bias of the QEC case is
\begin{align}
    B_{\mathrm{QEC}}=O(\Delta).
\end{align}
(In SM. Sec. S2, we provide an explicit form of the error-corrected state and the corresponding bias when stabilizer codes are employed.)

\section{VP with stabilizer state probe}
We show that VP can efficiently reduce a bias induced by the IIDP noise when we exploit a stabilizer state (without ancilla) as a quantum probe and adopt the same estimator defined in Eq. \eqref{mle}. 

First, we briefly review how the bias is modified under the application of VP. Detailed discussions can be found in Refs.~\cite{vp-PhysRevX.11.031057, vp-PhysRevX.11.041036, noi-yamamoto2021error,qem-kwon2024efficacy,supple}. To this end, we express the noisy state in its spectral decomposition as
\begin{align}
    \hat{\rho}_{\mathrm{e}}=\lambda \dyad{\psi_{\mathrm{e}}}+(1-\lambda)\hat{\sigma}_{\mathrm{e}},\label{errorstatediag}
\end{align}
where $\lambda$ denotes the largest eigenvalue, referred to as the \textit{dominant eigenvalue}, and $\ket{\psi_{\mathrm{e}}}$ is the corresponding eigenvector, referred to as the \textit{dominant eigenvector}. $\hat{\sigma}_{\mathrm{e}}$ is a density matrix supported on the subspace orthogonal to $\ket{\psi_{\mathrm{e}}}$. When the noisy process is characterized by a noise strength $\Delta$, all the terms in Eq.~\eqref{errorstatediag} depend implicitly on $\Delta$. In the noiseless limit $\Delta \to 0$, the error state $\hat{\rho}_{\mathrm{e}}$ reduces to the ideal state $\dyad{\psi}$. Consequently, the dominant eigenvalue approaches $1$ and the dominant eigenvector converges to the ideal state, i.e.,
\begin{align}
    \lim_{\Delta \to 0} \lambda =1,~~\lim_{\Delta \to 0}\ket{\psi_{\mathrm{e}}}=\ket{\psi}.
\end{align}
We now introduce the effect of VP through the purified error state
\begin{align}
    \frac{\hat{\rho}^{n}_{\mathrm{e}}}{\Tr\left[\hat{\rho}^{n}_{\mathrm{e}}\right]}=\frac{\lambda^{n}\dyad{\psi_{\mathrm{e}}}+(1-\lambda)^{n}\hat{\sigma}_{\mathrm{e}}}{\lambda^{n}+(1-\lambda)^{n}\mathrm{Tr}[\hat{\sigma}_{\mathrm{e}}^{n}]}.
\end{align}
In the small-noise regime $\Delta \ll 1$, we have $\lambda = O(1)$ and $1-\lambda = O(\Delta)$, which implies $\lambda^{n} \gg (1-\lambda)^{n}=O(\Delta^{n})$. As a result, the contribution of the dominant eigenvector $\ket{\psi_{\mathrm{e}}}$ is amplified, while the contribution from $\hat{\sigma}_{\mathrm{e}}$ is exponentially suppressed as $O(\Delta^{n})$. Therefore, provided that the dominant eigenvector $\ket{\psi_{\mathrm{e}}}$ is sufficiently close to the ideal state $\ket{\psi}$, purification enables an effectively recovery of the ideal state. However, implementing purification generally requires prior knowledge of the noisy state $\hat{\rho}_{\mathrm{e}}$, which motivates the use of VP that circumvent this requirement.

Rather than explicitly reconstructing the purified state, VP enables obtaining the mean of measurement outcomes $\bar{A}_{\mathrm{mit}}$ whose average over all possible measurement outcomes satisfies
\begin{align}
    \left<\bar{A}_{\mathrm{mit}}\right>=\frac{\Tr\left[\hat{A}\hat{\rho}^{n}_{\mathrm{e}}\right]}{\Tr\left[\hat{\rho}^{n}_{\mathrm{e}}\right]}:=\mu_{\mathrm{mit}}(\phi,\Delta) =\langle \psi_{\mathrm{e}} \vert \hat{A} \vert \psi_{\mathrm{e}} \rangle + O(\Delta^{n})
\end{align}
\cite{qem-kwon2024efficacy,supple}, without requiring prior knowledge of the error state. Detailed analysis and the implementation of VP can be found in Refs. \cite{vp-PhysRevX.11.031057, vp-PhysRevX.11.041036}. Similar to the noisy case, the theoretical reference function $\mu_{\mathrm{mit}}(\phi,\Delta)$ cannot be fully characterized due to the lack of complete knowledge of the noisy state $\hat{\rho}_{\mathrm{e}}$. We therefore employ the estimator
\begin{align}
    \phi^{\mathrm{est}}_{\mathrm{mit}}:= \mu^{-1}(\bar{A}_{\mathrm{mit}}),\label{estmit}
\end{align}
whose corresponding bias is given by
\begin{align}
    B_{\mathrm{mit}}:=\left<\phi^{\mathrm{est}}_{\mathrm{mit}}\right>-\phi = \mu^{-1}(\mu_{\mathrm{mit}}(\phi,\Delta))-\mu^{-1}(\mu(\phi)). \label{biasmit}
\end{align}
(A detailed derivation and the explicit expression of Eq. \eqref{biasmit} are provided in the Methods Sec. C.)
Eq.~\eqref{biasmit} shows that the bias originates from the discrepancy between the mitigated and ideal expectation values 
\begin{align}
    \mu_{\mathrm{mit}} - \mu = \langle \psi_{\mathrm{e}} \vert \hat{A} \vert \psi_{\mathrm{e}} \rangle - \langle \psi \vert \hat{A} \vert \psi \rangle + O(\Delta^{n}).
\end{align}
Hence, provided that the dominant eigenvector $\ket{\psi_{\mathrm{e}}}$ remains sufficiently close to the ideal state $\ket{\psi}$, the bias can be systematically reduced.

In this work, we focus on $n=2$, as it sufficiently demonstrates the efficacy of VP. We employ stabilizer states as the probe \cite{stb-nielsen2010quantum}. Specifically, we consider an $N$-qubit stabilizer state $\ket{\psi_{s0}}$, defined by its stabilizer group $G_{\mathbf{s}}$ with $\lvert G_{\mathbf{s}} \rvert = 2^{N}$, and consider $G_{\mathbf{s}}$ that does not contain single-qubit Pauli operators, i.e.,
\begin{align}
    \hat{X}^{(i)},\hat{Y}^{(i)},\hat{Z}^{(i)}  \not \in G_{\mathbf{s}} \forall i.
\end{align}
A representative example satisfying this condition is the GHZ state. Let us denote the corresponding signal state as $\ket{\psi_{s}} := \hat{U}(\phi)\ket{\psi_{s0}}$. In this setting, it is straightforward to verify that the error states $\hat{X}^{(i)}\ket{\psi_{s}}$, $\hat{Y}^{(i)}\ket{\psi_{s}}$, $\hat{Z}^{(i)}\ket{\psi_{s}}$, $\hat{U}(\phi)\hat{X}^{(i)}\ket{\psi_{s0}}$, and $\hat{U}(\phi)\hat{Y}^{(i)}\ket{\psi_{s0}}$, which constitute the first-order $O(\Delta)$ error components of the noisy state $\hat{\rho}_{\mathrm{e}}$, are all orthogonal to the ideal signal state $\ket{\psi_{s}}$ for every $1 \leq i \leq N$, independently of the value of $\phi$. Therefore, all first-order error processes involving the application of a single Pauli operator—either before or after the signal imprinting—produce states orthogonal to the ideal signal state. As a consequence, the dominant eigenvector of the noisy state satisfies
\begin{align}
    \dyad{\psi_{\mathrm{e}}}=\dyad{\psi_{s}}+O(\Delta^{2}),
\end{align}
indicating that deviations of the dominant eigenvector from the ideal signal state arise only at second order in the noise strength. As a consequence, bias of VP case is given by
\begin{align}
    B_{\mathrm{mit}}=O(\Delta^{2}).
\end{align}

This construction admits an interpretation analogous to that of QEC. We define the following subgroup of $G_{\mathbf{s}}$:
\begin{align}
    G_{\mathbf{s}^{[c]}} := \{\hat{S}\vert [\hat{S},\hat{Z}_{i}]=0~\forall ~1\leq i\leq N, \mathrm{ where } \hat{S} \in G_{\mathbf{s}} \}. \label{commutingstabilizer}
\end{align}
We denote by $\mathcal{D}$ the subspace stabilized by $G_{\mathbf{s}^{[c]}}$, and refer to it as the \textit{ideal space} or \textit{code space} generated by this subgroup. Eq. \eqref{commutingstabilizer} implies that every element of $G_{\mathbf{s}^{[c]}}$ commutes with the signal unitary $\hat{U}(\phi)$. Consequently, the ideal signal state $\ket{\psi_{s}}$ resides in the code space $\mathcal{D}$. Moreover, although all elements of $G_{\mathbf{s}^{[c]}}$ commute with the single Pauli $Z$ operators, the single Pauli $Z$ operators are not elements of $G_{\mathbf{s}^{[c]}}$. As a result, each single Pauli $Z$ operators acts as a logical Pauli $Z$ operator on the code space. This implies that the states $\ket{\psi_{s}}$ and $\hat{Z}\ket{\psi_{s}}$ cannot be distinguished by syndrome measurements, rendering a single Pauli $Z$ error operationally indistinguishable from the signal. However, it is worth emphasizing that although $\hat{Z}\ket{\psi_{s}}$ remains within the code space $\mathcal{D}$, it is still orthogonal to $\ket{\psi_{s}}$.  Therefore, while single Pauli
$Z$ errors cannot be detected via syndrome measurement, the bias induced by such errors can nevertheless be mitigated through the application of VP. (See Fig. \ref{fig:errorclass}.)

As a consequence, while QEC cannot correct the single Pauli $Z$ errors (because it is the logical error in the code space) resulting in a bias $B_{\mathrm{QEC}}=O(\Delta)$, VP can mitigate the errors, leading to a corresponding bias $B_{\mathrm{mit}}=O(\Delta^{2})$. This result implies the potential outperformance of VP compared to QEC in certain scenarios. Furthermore, in SM Sec. S2., we demonstrate that for a Pauli operator $\hat{A}$ satisfying the condition that there exists an $i$ such that $\{\hat{A},\hat{Z}_{i}\}=0$, in the presence of the local depolarizing with $p_{x}=p_{y}=p_{z}=\Delta/4$ or the local dephasing noise with $p_{z}=\Delta/2$, the bias of the QEC case scales as $B_{\mathrm{QEC}}=\Theta(\Delta)$ except for values of $\phi$ that satisfy $\mu(\phi)=0$. In contrast, the bias of VP exploiting stabilizer state as a quantum probe scales as $B_{\mathrm{mit}}=O(\Delta^{2})$. In addition, we also show that the bias of VP case can be further reduced to $O(\Delta^{n \geq 2})$. See SM. Sec. 3 for the derivation.

\section{Statistical Error Analysis}
Lastly, we compare the statistical errors in the noisy (including QEC) scenario and in the VP approach. For the noisy or QEC-assisted case—when the noise cannot be fully corrected—and using the estimator defined in Eq.~\eqref{esterror}, the statistical error is given by
\begin{align}
    \left<\left(\phi^{\mathrm{est}}_{\mathrm{e}}\right)^{2}\right>-\left<\phi^{\mathrm{est}}_{\mathrm{e}}\right>^{2}=\left(\frac{1}{\pdv{\mu(\phi)}{\phi}\vert_{\phi=\phi_{\mathrm{e}}}}\right)^{2}\sigma_{\mathrm{e}}(\phi)^{2}, \label{staterrore}
\end{align}
where $\phi_{\mathrm{e}}$ satisfies $\mu(\phi_{\mathrm{e}})=\mu_{\mathrm{e}}(\phi)$ and 
\begin{align}
    \sigma_{\mathrm{e}}(\phi)^{2}:= \frac{\left<\bar{A}^{2}_{\mathrm{e}}\right>-\left<\bar{A}_{\mathrm{e}}\right>^{2}}{M} =\frac{1-\Tr\left[\hat{A}\hat{\rho}_{\mathrm{e}}(\phi)\right]^{2}}{M}.
\end{align} 
Next, for the VP case, using the estimator defined in Eq.~\eqref{estmit}, the statistical error takes the form
\begin{align}
    \left<\left(\phi^{\mathrm{est}}_{\mathrm{mit}}\right)^{2}\right>-\left<\phi^{\mathrm{est}}_{\mathrm{mit}}\right>^{2}=\left(\frac{1}{\pdv{\mu(\phi)}{\phi}\vert_{\phi=\phi_{\mathrm{mit}}}}\right)^{2}\sigma_{\mathrm{mit}}(\phi)^{2}
\end{align}
where $\phi_{\mathrm{mit}}$ satisfies $\mu(\phi_{\mathrm{mit}})=\mu_{\mathrm{mit}}(\phi)$ and $\sigma_{\mathrm{mit}}^{2}$ is \cite{noi-yamamoto2021error,vp-PhysRevX.11.031057,vp-PhysRevX.11.041036,qem-kwon2024efficacy},
\begin{align}
    \begin{split}
    &\sigma_{\mathrm{mit}}^{2} := \frac{2n}{M}\bigg[\frac{1-\Tr\left[\hat{A}\hat{\rho}^{n}_{\mathrm{e}}\right]^{2}}{\Tr\left[\hat{\rho}^{n}_{\mathrm{e}}\right]^{2}} +\frac{\Tr\left[\hat{A}\hat{\rho}^{n}_{\mathrm{e}}\right]^{2} (1-\Tr\left[\hat{\rho}^{n}_{\mathrm{e}}\right]^{2})}{\Tr\left[\hat{\rho}^{n}_{\mathrm{e}}\right]^{4}} \bigg].
    \end{split}\label{staterrormit}
\end{align}
Notably, the following quantities appearing in Eq.~\eqref{staterrore} can be approximated as
\begin{align}
    \begin{split}
    &\Tr\left[\hat{\rho}^{n}_{\mathrm{e}}\right]^{2} =  \lambda^{2n}+O(\Delta^{2n}),\\
    &\Tr\left[\hat{A}\hat{\rho}^{n}_{\mathrm{e}}\right]^{2} = \lambda^{2n}\langle \psi_{\mathrm{e}}\vert \hat{A} \vert \psi_{\mathrm{e}}\rangle +O(\Delta^{2n}).
    \end{split}
\end{align}
Substituting these approximations into Eq.~\eqref{staterrormit}, we find that the statistical error of the mitigated estimator scales as
\begin{align}
    \sigma_{\mathrm{mit}}^{2} \propto \frac{2n}{\lambda^{2n}},
\end{align}
up to multiplicative factors of order unity.
This implies that the statistical error increases exponentially with the purification order $n$, reflecting the well-known trade-off between bias suppression and an increase in variance inherent to the VP approach \cite{qem-takagi2022fundamental, qem-PhysRevLett.131.210601, qem-PhysRevLett.131.210602, vp-PhysRevX.11.031057, vp-PhysRevX.11.041036}. Nevertheless, such an increase in statistical error does not necessarily preclude an overall improvement in the total mean squared error.
In particular, for $n=2$, the induced statistical error remains modest, such that the reduction in bias leads to a net decrease in the total mean squared error. As shown in Eqs.~\eqref{staterrore} and \eqref{staterrormit}, the statistical errors in both the noisy (or QEC-assisted) case and the VP case scale inversely with the sample number $M$. In contrast, the bias terms in Eqs.~\eqref{biasgeneral} and \eqref{biasmit} are independent of $M$. Consequently, in the regime of sufficiently large sample size, the bias becomes the dominant contribution to the mean squared error, as defined in Eq.~\eqref{MSE}. In this regime, the bias reduction achieved by VP can outweigh the accompanying increase in statistical error, leading to a net enhancement in estimation performance. See the numerical results presented in the next section and in Sec.~S4 of the SM.

\section{Numerical results}
\begin{figure}[t]
    \centering
    \includegraphics[width=0.99\linewidth]{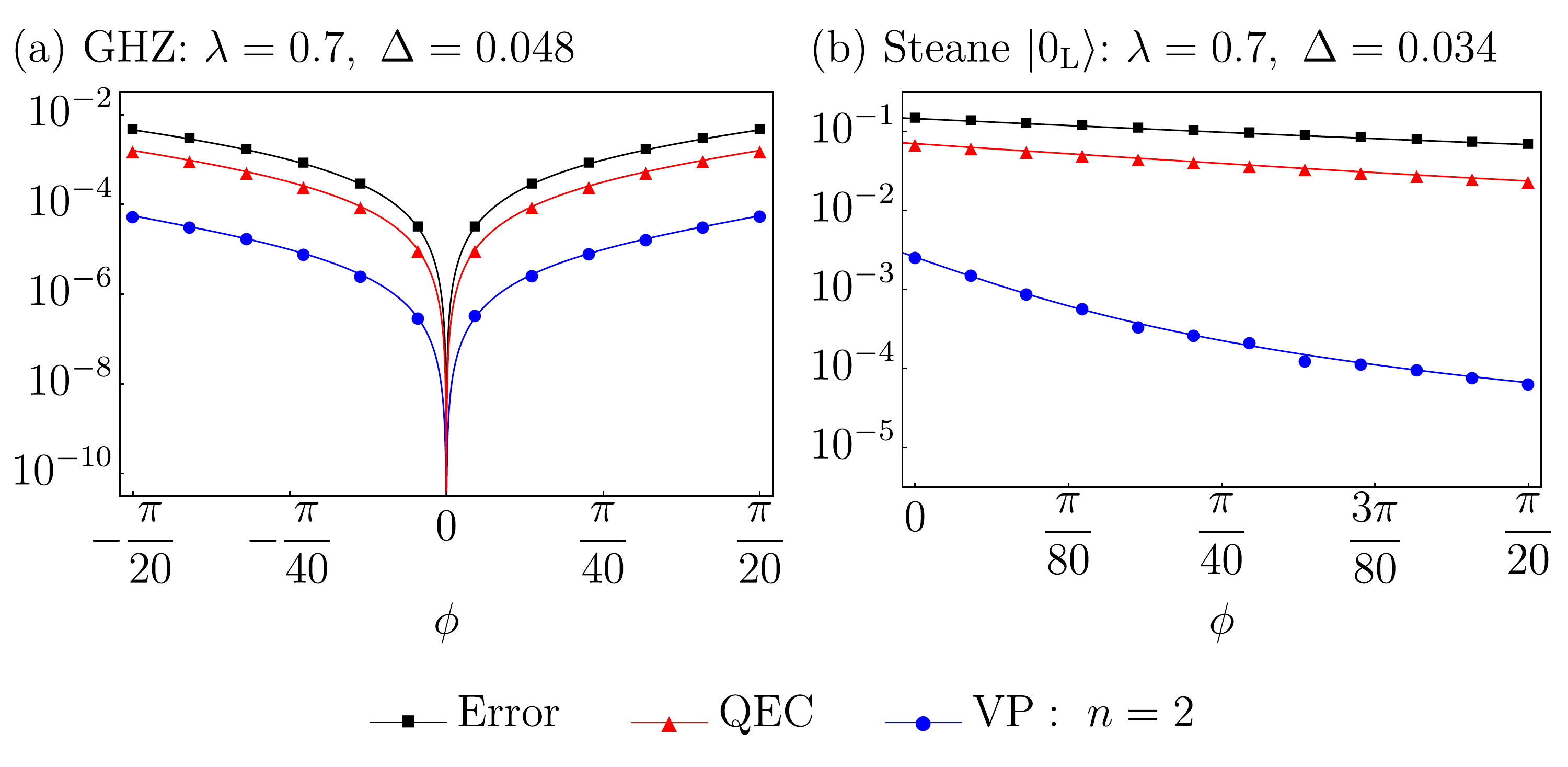}
    \caption{(a)-(b) 
    The square of biases (shown on a logarithmic scale) in the estimation of $\phi \in \left[-\frac{\pi}{20}, \frac{\pi}{20}\right]$, obtained using (a) $5$-qubit GHZ state (b) logical $0$ state of the $7$-qubit Steane code as quantum probes, in the presence of local depolarizing noise with the noise strengths that render the largest eigenvalue as $\lambda=0.7$. Solid lines indicate the theoretical values of the square of biases, while circles, triangles, and squares represent the total estimation error, based on the maximum likelihood estimation obtained with $M = 10^9$ samples. The ``Error'' case corresponds to the scenario where none of the schemes are applied. We emphasize that VP enables a substantial improvement in estimation performance—namely, a significantly reduced total estimation error in the large sample limit, since in this regime, the bias becomes the dominant contribution to the total estimation error.
    } 
    \label{fig:GHZPauli}
\end{figure}

For the numerical simulations, we consider two distinct quantum probes: 5-qubit GHZ state, and the logical $\ket{0}$ state of the $7$-qubit Steane code, as the representative example of stabilizer states \cite{stb-nielsen2010quantum}. We assume that both states are subjected to local depolarizing noise characterized by $p_{x} = p_{y} = p_{z} = \Delta/4$. For the QEC scenarios, the same quantum probes are utilized with the aid of an ancillary system and we consider stabilizer code. We explore noise strength that corresponds to the dominant eigenvalue of the error state at $0.7$. The results for each quantum probe are illustrated in Fig. \ref{fig:GHZPauli}, (a), and (b). In both cases, we observe that VP substantially reduces bias induced by the local depolarizing noise, compared to QEC. Notably, in the case of the GHZ state, all biases vanish at $\phi=0$ since the expectation values of $\hat{A}$ coincidentally become identical at this point. Furthermore, for the GHZ state, the expectation values of $\hat{A}$ are sine functions, which allow for the estimation of $\phi$ for both positive and negative values. In contrast, for the $7$-qubit case, the expectation values follow an even functional form, resulting in the same distribution for $\phi$ and $-\phi$. Consequently, in this scenario, $\phi$ can only be estimated within the positive (or negative) domain. The detailed settings for each scenario can be found in SM. Sec. S4.

\section{Discussion}
We have analyzed the performance of VP and QEC in the canonical phase estimation scenario, employing Pauli measurements for estimation in the presence of Pauli noise. We show that VP can effectively mitigate bias induced by indistinguishable noise, which even an ideal QEC process cannot fully correct. We then demonstrate that combining VP with a stabilizer state quantum probe efficiently suppresses bias errors outperforming QEC. Numerical simulations show that VP, employing a probe in 5-qubit GHZ state or encoded in 7-qubit Steane code, significantly reduces bias compared to QEC with the same probe states. This result highlights that VP can complement, or serve as a robust alternative to QEC by effectively addressing indistinguishable errors that QEC alone cannot mitigate. We believe that our work provides a practical and effective strategy for enhancing accuracy in noisy quantum metrology, particularly when indistinguishable errors pose significant challenges.

Finally, our results motivate exploring whether error-mitigation techniques beyond VP can also outperform QEC in quantum metrology. In particular, it would be interesting to investigate whether \textit{virtual channel purification} \cite{discusvcp-PRXQuantum.6.020325}, which leverages purification at the channel level, can similarly suppress bias and yield a metrological advantage, analogous to the state-level purification realized by VP. Beyond purification-based approaches, while \textit{zero-noise extrapolation} is a natural candidate \cite{discuszero-PhysRevLett.119.180509, discuszero-PhysRevX.7.021050}, recent works—although considered in settings different from that of this manuscript—suggest that directly learning or characterizing the noise may require fewer samples and offer superior performance \cite{discus-ijaz2024more}. Moreover, \textit{probabilistic error cancellation} is unlikely to be effective in the presence of unknown noise, as it fundamentally relies on accurate noise characterization \cite{discuszero-PhysRevLett.119.180509}. It is therefore worthwhile to examine whether other error-mitigation techniques, such as \textit{quantum subspace expansion} \cite{discussubex-mcclean2020decoding, discussubex-PhysRevA.95.042308, discussubex-PhysRevX.10.011004, discussubex-PhysRevLett.129.020502}, \textit{symmetry verification} \cite{discussymetry-PhysRevA.98.062339, discussymetry-PhysRevLett.122.180501, discussymetry-jcd6-lft3}, or \textit{error filtration} \cite{discusef-PhysRevLett.131.190601}—can suppress bias and improve estimation performance compared to QEC.

\section{Methods}
\subsection{MLE}
We introduce the \textit{maximum likelihood estimator} (MLE), a statistical method used to estimate parameters based on measurement outcomes. The conditional joint probability that we obtain measurement outcomes $\vec{A}=(A_{1},A_{2},\cdots , A_{M})$ when the signal parameter is a given value $\phi$, denoted as $P(\vec{A}\vert\phi)$, is called the \textit{likelihood function}. Here, $A_{i}$ is the $i$th measurement outcome. When each measurement outcome follows an independent identical distribution, the likelihood function is given by
\begin{align}
    P(\vec{A}\vert\phi) =\prod_{i=1}^{M}P(A_{i}\vert \phi).
\end{align}
The \textit{maximum likelihood estimation} adopts the value that maximizes the likelihood function as an estimator of $\phi$, i.e., 
\begin{align}
    \phi^{\mathrm{est}}=\arg \max_{\theta} P(\vec{A}\vert\theta).
\end{align}

Next, let us inspect the MLE when the eigenvalues of an observable $\hat{A}$ are $1$ and $-1$ which corresponds to the Pauli measurement that we mainly focus on in this paper. We assume that the measurement outcome $A_{i}$ follows independent identical binomial distribution for all $i=1,2,\cdots,M$. In this case, since all the measurement outcomes follow the binomial distribution, 
\begin{align}
    P(\vec{A}\vert \theta)=\prod_{i=1}^{M}P(A_{i}\vert \theta)=p(\theta)^{n_{1}}(1-p(\theta))^{M-n_{1}},
\end{align}
where $p(\theta) \equiv \frac{1}{2}\left(1+\Tr\left[\hat{A}\hat{\rho}(\theta)\right]\right)$ is the probability that the measurement outcome is $1$, and $n_{1} \equiv \sum_{k=1}^{M}\left(\frac{A_{k}+1}{2}\right)$ is the number of $1$ among $M$ number of measurement outcomes. To find the MLE, let us find $\theta$ that maximizes $P(\vec{A}\vert \theta)$ using the \textit{log-likelihood function} which we denote as $L(\theta)\equiv \ln{P(\vec{A}\vert \theta)}$. At the parameter value that maximizes the likelihood function (equivalently, the value maximizes the log-likelihood function), the derivative of $L(\theta)$ becomes $0$:
\begin{align}
    \pdv{}{\theta}L(\theta)\bigg\vert_{\theta=\phi^{\mathrm{est}}}=\pdv{p(\theta)}{\theta}\left[\frac{n_{1}-Mp(\theta)}{p(\theta)(1-p(\theta))}\right] \bigg\vert_{\theta=\phi^{\mathrm{est}}}=0.
\end{align}
Therefore, when $\pdv{p(\theta)}{\theta}\neq 0$, $\phi^{\mathrm{est}}$ satisfies
\begin{align}
    p(\phi^{\mathrm{est}})=\frac{1}{2}\left(1+\mu(\phi^{\mathrm{est}})\right)=\frac{n_{1}}{M}=\frac{1}{2}\left(1+\bar{A}\right),
\end{align}
which results in 
\begin{align}
    \phi^{\mathrm{est}}=\mu^{-1}(\bar{A}), \label{method:mle22}
\end{align}
where $\mu(\phi)\equiv \Tr\left[\hat{A}\hat{\rho}(\phi)\right]$, $\bar{A}\equiv \sum_{i=1}^{M} A_{i}/M$, and $\mu^{-1}$ is the inverse function of $\mu(\phi)$. To find the asymptotic behavior of $\phi^{\mathrm{est}}$ in the large $M$ limit, let us consider its Taylor's exapnsion
\begin{align}
    \phi^{\mathrm{est}} =\mu^{-1}(\mu(\phi))+\sum_{k=1}^{\infty}M_{k}(\bar{A}-\mu(\phi))^{k}, \label{method:estimatorofphi}
\end{align}
where $M_{k}\equiv \frac{1}{k!}\pdv{^{k}}{x^{k}}\mu^{-1}(x)\big\vert_{x=\mu(\phi)}$. In the large sample limit, $\bar{A}$ follows the normal distribution $\mathcal{N}(\mu,\sigma)$ (because of the central limit theorem,) where $\mu(\phi) = \left<\bar{A}\right>=\Tr\left[\hat{A}\hat{\rho}(\phi)\right]$ and $\sigma(\phi)^{2} \equiv \left<\bar{A}^{2}\right>-\left<\bar{A}\right>^{2}=\frac{1-\Tr\left[\hat{A}\hat{\rho}(\phi)\right]^{2}}{M}$. According to the formulas of central moments of normal distributions \cite{supp-edition2002probability} which are
\begin{align}
    &\left<(\bar{A}-\mu)\right>^{2n}=\frac{(2n)!\sigma^{2n}}{n!2^{n}}, ~~\left<(\bar{A}-\mu)\right>^{2n+1}=0, \label{method:oddapprox}
\end{align}
the expectation value of the estimator is
\begin{align}
    \begin{split}
    &\left<\phi^{\mathrm{est}}\right> = \mu^{-1}(\mu(\phi))+\sum_{k=1}^{\infty}f_{2k}\left<(\bar{A}-\mu(\phi))^{2k}\right>\\
    &=\phi+\sum_{k=1}^{\infty}f_{2k}\frac{(2k)!}{k!2^{k}}\sigma(\phi)^{2k}. 
    \end{split}
    \label{method:expecMLE}
\end{align}
Lastly since $\sigma \to 0$ in the large sample limit $M \to \infty$,
\begin{align}
    \lim_{M\to \infty}\left<\phi^{\mathrm{est}}\right>=\phi.
\end{align}

Using Eqs. \eqref{method:estimatorofphi} and \eqref{method:oddapprox}, one can find that the corresponding statistical error is
\begin{align}
    &\left<\left(\phi^{\mathrm{est}}\right)^{2}\right>-\left<\left(\phi^{\mathrm{est}}\right)\right>^{2}= \left(\frac{1}{\pdv{\mu(\phi)}{\phi}}\right)^{2}\sigma(\phi)^{2}. \label{method:estvari}
\end{align}

\subsection{Noisy case}
When the noise characterization is intractable, a practical choice of an estimator would be 
\begin{align}
    \phi^{\mathrm{est}}_{\mathrm{e}}\equiv \mu^{-1}(\bar{A}_{\mathrm{e}}),     \label{method:esterror}
\end{align}
since $\mu(\phi)$ is known. In analogous to Eq. \eqref{method:expecMLE}, one can find that the expectation value of such an estimator in the large sample limit is 
\begin{align}
    \left<\phi^{\mathrm{est}}_{\mathrm{e}}\right>=\mu^{-1}(\mu_{\mathrm{e}}(\phi)),
\end{align}
where $\mu_{\mathrm{e}}\equiv \Tr[\hat{A}\hat{\rho}_{\mathrm{e}}]$.
Therefore the bias of $\phi^{\mathrm{est}}_{\mathrm{e}}$ is
\begin{align}
    \begin{split}
    &B_{\mathrm{e}}:= \left<\phi^{\mathrm{est}}_{\mathrm{e}}\right>-\phi=\mu^{-1}(\mu_{\mathrm{e}}(\phi))-\mu^{-1}(\mu(\phi))\\
    &=\sum_{k=1}^{\infty} M_{k}\left(\mu_{\mathrm{e}}-\mu\right)^{k}=\frac{\mu_{\mathrm{e}}-\mu}{\pdv{\mu}{\phi}}+O\left((\mu_{\mathrm{e}}-\mu)^{2}\right).
    \end{split}
\end{align}
where $M_{k}\equiv \frac{1}{k!}\pdv{^{k}}{x^{k}}\mu^{-1}(x)\big\vert_{x=\mu}$. Next, let us express the bias $B_{\mathrm{e}}$ in terms of the noise strength $\Delta$. First, let us inspect $\mu_{\mathrm{e}}(\phi)-\mu(\phi)$:
\begin{align}
    \begin{split}
    &\mu_{\mathrm{e}}(\phi)-\mu(\phi)\\
    &=\lambda \left[\langle \psi_{\mathrm{e}}  \vert \hat{A} \vert \psi_{\mathrm{e}} \rangle - \langle \psi \vert \hat{A} \vert \psi \rangle \right]+(1-\lambda)\left[\Tr[\hat{A}\hat{\sigma}_{\mathrm{e}}] - \langle \psi \vert \hat{A} \vert \psi \rangle\right]\\
    &=\left[1 - \sum_{k=1}^{\infty}\lambda_{k}\Delta^{k}\right]\left[\sum_{k=1}^{\infty}a_{k}\Delta^{k}\right]+\left[\sum_{k=1}^{\infty}\lambda_{k}\Delta^{k}\right]\left[\sum_{k=0}^{\infty}b_{k}\Delta^{k}\right]\\
    &=\sum_{k=1}^{\infty}\left[a_{k} + \sum_{l=1}^{k}\lambda_{l}a_{k-l} - \sum_{l=1}^{k}\lambda_{l}b_{k-l}\right]\Delta^{k}= \sum_{k=1}^{\infty}f_{k}\Delta^{k}, \label{mueorder}
    \end{split}
\end{align}
where the relevant terms $\lambda_{k}$, $a_{k}$, $b_{k}$ and, $f_{k}$ are defined as
\begin{align}
    &1-\lambda = \sum_{k=0}^{\infty}\lambda_{k}(\phi)\Delta^{k}, \label{lambdaterm}\\
    &\langle \psi_{\mathrm{e}} \vert \hat{A} \vert \psi_{\mathrm{e}} \rangle - \langle \psi \vert \hat{A} \vert \psi \rangle = \sum_{k=0}^{\infty}a_{k}(\phi)\Delta^{k}, \label{dominant}\\ 
    &\Tr[\hat{A}\hat{\sigma}_{\mathrm{e}}] - \langle \psi \vert \hat{A} \vert \psi \rangle = \sum_{k=0}^{\infty}b_{k}(\phi)\Delta^{k}, \label{tailterm}\\
    &f_{k}(\phi)\equiv a_{k}(\phi) - \sum_{l=0}^{k}\lambda_{l}(\phi)a_{k-l}(\phi) + \sum_{l=0}^{k}\lambda_{l}(\phi)b_{k-l}(\phi).
\end{align}
Finally, the bias can be expressed as
\begin{align}
    B_{\mathrm{e}}=\frac{1}{\frac{\mu(\phi)}{d\phi}}f_{1}\Delta+O(\Delta^{2}) = O(\Delta). \label{method:biaserror}
\end{align}

Next, let us find the statistical error $\left<\left(\phi^{\mathrm{est}}_{\mathrm{e}}\right)^{2}\right>-\left<\phi^{\mathrm{est}}_{\mathrm{e}}\right>^{2}$. Since $\phi^{\mathrm{est}}_{\mathrm{e}}=\mu^{-1}(\bar{A}_{\mathrm{e}})$, can be expressed as
\begin{align}
    \phi^{\mathrm{est}}_{\mathrm{e}}=\mu^{-1}(\bar{A}_{\mathrm{e}})=\sum_{k=0}^{\infty}M_{\mathrm{e},k}(\bar{A}_{\mathrm{e}}-\mu_{\mathrm{e}}(\phi))^{k}.
\end{align}
Here $M_{\mathrm{e},k}\equiv \frac{1}{k!}\pdv{^{k}}{x^{k}}\mu^{-1}(x)\big\vert_{x=\mu_{\mathrm{e}}}$. Therefore, according to Eq. \eqref{method:oddapprox},
\begin{align}
    \begin{split}
    &\left<\left(\phi^{\mathrm{est}}_{\mathrm{e}}\right)^{2}\right>\\
    &=\left(M_{\mathrm{e},0}\right)^{2}+\left(2M_{\mathrm{e},0}M_{\mathrm{e},2}+M_{\mathrm{e},1}M_{\mathrm{e},1}\right)\sigma_{\mathrm{e}}(\phi)^{2}+O\left(\frac{1}{\left(M\right)^{4}}\right),\\
    &\left<\phi^{\mathrm{est}}_{\mathrm{e}}\right>^{2}=\left(M_{\mathrm{e},0}\right)^{2}+\left(2M_{\mathrm{e},0}M_{\mathrm{e},2}\right)\sigma_{\mathrm{e}}(\phi)^{2}+O\left(\frac{1}{\left(M\right)^{4}}\right),
    \end{split}
\end{align}
where $\sigma_{\mathrm{e}}(\phi)^{2}\equiv \frac{\left<\bar{A}^{2}_{\mathrm{e}}\right>-\left<\bar{A}_{\mathrm{e}}\right>^{2}}{M} =\frac{1-\Tr\left[\hat{A}\hat{\rho}_{\mathrm{e}}(\phi)\right]^{2}}{M}$. As a result, in the large sample limit $M \to \infty$, the statistical error is 
\begin{align}
    \left<\left(\phi^{\mathrm{est}}_{\mathrm{e}}\right)^{2}\right>-\left<\phi^{\mathrm{est}}_{\mathrm{e}}\right>^{2}=\left(\frac{1}{\pdv{\mu(\phi)}{\phi}\vert_{\phi=\phi_{\mathrm{e}}}}\right)^{2}\sigma_{\mathrm{e}}(\phi)^{2}, \label{method:staterrore}
\end{align}
where $\phi_{\mathrm{e}}$ satisfies $\mu(\phi_{\mathrm{e}})=\mu_{\mathrm{e}}(\phi)$.

Lastly, we express the statistical error in Eq. \eqref{method:staterrore} in the asymptotic limit $\Delta \to 0$. In this regime, the dominant eigenvalue satisfies $\lambda \gg (1-\lambda)$, so that the contribution of the subleading component of the state becomes perturbative. Accordingly, we obtain 
\begin{align}
    \begin{split}
    &M\sigma_{\mathrm{e}}(\phi)^{2}=1-\Tr\left[\hat{A}\hat{\rho}_{\mathrm{e}}(\phi)\right]^{2}\\
    &=1-\lambda^{2}\langle \psi_{\mathrm{e}} \vert \hat{A} \vert \psi_{\mathrm{e}}\rangle^{2}+O((1-\lambda)).
    \end{split}
\end{align}
Substituting this expression into Eq. \eqref{method:staterrore}, we find that the statistical error of the estimator is approximately given by
\begin{align}
    \begin{split}
    &\left<\left(\phi^{\mathrm{est}}_{\mathrm{e}}\right)^{2}\right>-\left<\phi^{\mathrm{est}}_{\mathrm{e}}\right>^{2} \\
    &= \left(\frac{1}{\pdv{\mu(\phi)}{\phi}\vert_{\phi=\phi_{\mathrm{e}}}}\right)^{2}\left(\frac{1-\lambda^{2}\langle \psi_{\mathrm{e}} \vert \hat{A} \vert \psi_{\mathrm{e}}\rangle^{2}}{M}\right)+O((1-\lambda))
    \end{split}
\end{align}
which explicitly highlights how the estimator variance is governed, in the asymptotic regime.

\subsection{VP case}
Similar to the noisy case, one can consider the following estimator for the mitigation case:
\begin{align}
    \phi^{\mathrm{est}}_{\mathrm{mit}}\equiv \mu^{-1}(\bar{A}_{\mathrm{mit}}), \label{method:estmit}
\end{align}
whose expectation value is
\begin{align}
    \left<\phi^{\mathrm{est}}_{\mathrm{mit}}\right> = \mu^{-1}(\mu_{\mathrm{mit}}(\phi)).
\end{align}
The bias of the corresponding estimator is
\begin{align}
    \begin{split}
    &B_{\mathrm{mit}}:= \left<\phi^{\mathrm{est}}_{\mathrm{mit}}\right>-\phi=\mu^{-1}(\mu_{\mathrm{mit}}(\phi))-\mu^{-1}(\mu(\phi)) \\
    &= \sum_{k=1}^{\infty} M_{k}\left(\mu_{\mathrm{mit}}-\mu\right)^{k}=\frac{\mu_{\mathrm{mit}}-\mu}{\pdv{\mu}{\phi}}+O(\mu_{\mathrm{mit}}-\mu)^{2},
    \end{split}
\end{align}
where $M_{k}\equiv \frac{1}{k!}\pdv{^{k}}{x^{k}}\mu^{-1}(x)\big\vert_{x=\mu}$.
Next, let us inspect $\mu_{\mathrm{mit}}(\phi)-\mu(\phi)$. For the simpler expression, let us define the coefficient $\lambda'$ whose Taylor expansion is
\begin{align}
    \begin{split}
    &\lambda'=\frac{\lambda^{n}}{\lambda^{n}+(1-\lambda)^{n}\Tr[(\hat{\sigma}_{\mathrm{e}})^{n}]}\\
    &=\frac{1}{1+\left(\frac{1-\lambda}{\lambda}\right)^{n}\Tr[(\hat{\sigma}_{\mathrm{e}})^{n}]}\\
    &=1+\sum_{k=1}^{\infty}\left[-\left(\frac{1-\lambda}{\lambda}\right)^{n}\Tr[(\hat{\sigma}_{\mathrm{e}})^{n}]\right]^{k}\\
    &:= 1-\sum_{k=n}^{\infty}\lambda'_{k}\Delta^{k},
    \end{split}
\end{align}
$\mu_{\mathrm{mit}}(\phi)-\mu(\phi)$ is then
\begin{align}
    \begin{split}
    &\mu_{\mathrm{mit}}(\phi)-\mu(\phi)\\
    &=\lambda' \left[\langle \psi_{\mathrm{e}}  \vert \hat{A} \vert \psi_{\mathrm{e}} \rangle - \langle \psi \vert \hat{A} \vert \psi \rangle \right]\\
    &+(1-\lambda')\left[\Tr[\hat{A}(\hat{\sigma}_{\mathrm{e}})^{n}] - \langle \psi \vert \hat{A} \vert \psi \rangle\right]\\
    &=\left[1 - \sum_{k=n}^{\infty}\lambda'_{k}\Delta^{k}\right]\left[\sum_{k=1}^{\infty}a_{k}\Delta^{k}\right]+\sum_{k=n}^{\infty}\lambda'_{k}\left[\sum_{k=0}^{\infty}c_{k}\Delta^{k}\right]\\
    &=\sum_{k=1}^{n-1}a_{k}\Delta^{k}+ \sum_{k=n}^{\infty} \left[a_{k} + \sum_{l=n}^{k}\lambda'_{l}a_{k-l} - \sum_{l=n}^{k}\lambda'_{l}c_{k-l}\right] \Delta^{k}\\
    &:= \sum_{k=1}^{n-1}a_{k}\Delta^{k} + \sum_{k=n}^{\infty} g_{k}\Delta^{k} = \sum_{k=1}^{n-1}a_{k}\Delta^{k} + O(\Delta^{n}) \label{method:mumitorder}
    \end{split}
\end{align}
where the relevant term $c_{k}$ and $g_{k}$ are
\begin{align}
    &\Tr[\hat{A}(\hat{\sigma}_{\mathrm{e}})^{n}] - \langle \psi \vert \hat{A} \vert \psi \rangle = \sum_{k=0}^{\infty}c_{k}\Delta^{k}, \\
    &g_{k} = a_{k} + \sum_{l=n}^{k}\lambda'_{l}a_{k-l} - \sum_{l=n}^{k}\lambda'_{l}c_{k-l}.
\end{align}
Finally, the bias can be expressed as
\begin{align}
    B_{\mathrm{mit}}=\frac{1}{\frac{d\mu(\phi)}{d\phi}}\sum_{k=1}^{n-1}a_{k}\Delta^{k}+O(\Delta^{n}). \label{method:biasmit}
\end{align}

Next, let us inspect the statistical error $\left<\left(\phi^{\mathrm{est}}_{\mathrm{mit}}\right)^{2}\right>-\left<\phi^{\mathrm{est}}_{\mathrm{mit}}\right>^{2}$. Following the same techniques that we used for the derivation of Eq. \eqref{method:staterrore}, one can find that
\begin{align}
    \left<\left(\phi^{\mathrm{est}}_{\mathrm{mit}}\right)^{2}\right>-\left<\phi^{\mathrm{est}}_{\mathrm{mit}}\right>^{2}=\left(\frac{1}{\pdv{\mu(\phi)}{\phi}\vert_{\phi=\phi_{\mathrm{mit}}}}\right)^{2}\sigma_{\mathrm{mit}}(\phi)^{2}
\end{align}
where $\phi_{\mathrm{mit}}$ satisfies $\mu(\phi_{\mathrm{mit}})=\mu_{\mathrm{mit}}(\phi)$ and $\sigma_{\mathrm{mit}}^{2}$ is the variance of $\bar{A}_{\mathrm{mit}}$, which can be found in Refs. \cite{noi-yamamoto2021error,vp-PhysRevX.11.031057,vp-PhysRevX.11.041036,qem-kwon2024efficacy}, is
\begin{align}
    \begin{split}
    &\sigma_{\mathrm{mit}}^{2}\equiv \frac{\left<\bar{A}^{2}_{\mathrm{mit}}\right>-\left<\bar{A}_{\mathrm{mit}}\right>^{2}}{M}\\
    &=\frac{2n}{M}\bigg[\frac{1-\Tr\left[\hat{A}\hat{\rho}^{n}_{\mathrm{e}}\right]^{2}}{\Tr\left[\hat{\rho}^{n}_{\mathrm{e}}\right]^{2}} +\frac{\Tr\left[\hat{A}\hat{\rho}^{n}_{\mathrm{e}}\right]^{2} (1-\Tr\left[\hat{\rho}^{n}_{\mathrm{e}}\right]^{2})}{\Tr\left[\hat{\rho}^{n}_{\mathrm{e}}\right]^{4}} \bigg].
    \end{split}\label{method:staterrormit}
\end{align}
We further analyze the statistical error in Eq. \eqref{method:staterrormit} in the asymptotic limit $\Delta \to 0$ which implies $\lambda \gg (1-\lambda)$. In this regime, the terms in Eq. \eqref{method:staterrormit} can be expressed as (in terms of $\lambda$ and $(1-\lambda)$)
\begin{align}
    \begin{split}
    &\frac{1-\Tr\left[\hat{A}\hat{\rho}^{n}_{\mathrm{e}}\right]^{2}}{\Tr\left[\hat{\rho}^{n}_{\mathrm{e}}\right]^{2}} = \frac{1-\left(\lambda^{n}\langle \psi_{\mathrm{e}} \vert \hat{A} \vert \psi_{\mathrm{e}}\rangle+(1-\lambda)^{n}\mathrm{Tr}[\hat{A}(\hat{\sigma}_{\mathrm{e}})^{n}]\right)^{2}}{\left(\lambda^{n}+(1-\lambda)^{n}\mathrm{Tr}[(\hat{\sigma}_{\mathrm{e}})^{n}]\right)^{2}} \\
    &= \frac{1-\lambda^{2n}\langle \psi_{\mathrm{e}} \vert \hat{A} \vert \psi_{\mathrm{e}}\rangle^{2}}{\lambda^{2n}}+O((1-\lambda)),
    \end{split}
\end{align}
and
\begin{align}
    \begin{split}
    &\frac{\Tr\left[\hat{A}\hat{\rho}^{n}_{\mathrm{e}}\right]^{2} (1-\Tr\left[\hat{\rho}^{n}_{\mathrm{e}}\right]^{2})}{\Tr\left[\hat{\rho}^{n}_{\mathrm{e}}\right]^{4}}\\
    &=\frac{\left(\lambda^{n}\langle \psi_{\mathrm{e}} \vert \hat{A} \vert \psi_{\mathrm{e}}\rangle+(1-\lambda)^{n}\mathrm{Tr}[\hat{A}(\hat{\sigma}_{\mathrm{e}})^{n}]\right)^{2}}{\left(\lambda^{n}+(1-\lambda)^{n}\mathrm{Tr}[(\hat{\sigma}_{\mathrm{e}})^{n}]\right)^{4}}\\
    &\times \left(1-\left(\lambda^{n}+(1-\lambda)^{n}\mathrm{Tr}[(\hat{\sigma}_{\mathrm{e}})^{n}]\right)^{2}\right)=O((1-\lambda)).
    \end{split}
\end{align}
As a consequence, the statistical error in Eq. \eqref{method:staterrormit} can be approximated as
\begin{align}
    \begin{split}
    &\left<\left(\phi^{\mathrm{est}}_{\mathrm{mit}}\right)^{2}\right>-\left<\phi^{\mathrm{est}}_{\mathrm{mit}}\right>^{2}\\
    &=\left(\frac{1}{\pdv{\mu(\phi)}{\phi}\vert_{\phi=\phi_{\mathrm{mit}}}}\right)^{2}\frac{2n}{\lambda^{2n}}\left(\frac{1-\lambda^{2n}\langle \psi_{\mathrm{e}}\vert \hat{A} \vert \psi_{\mathrm{e}} \rangle }{M}\right)\\
    &+O((1-\lambda)).
    \end{split}
\end{align}
This expression explicitly shows the exponential amplification of statistical fluctuations with increasing $n$, governed by the factor $\lambda^{-2n}$.

\subsection{QEC}\label{method:secqec}
Suppose the IIDP noise defined in Eq. \eqref{IIDPnoise} with $p_{z} \neq 0$. We emphasize that for such a IIDP noise, $\{\hat{I}\}\cup \{\hat{Z}^{(j)}\}_{j=1}^{N} \subset \{\hat{E}_{i}\}_{i}$. To correct all the single Pauli $Z$ errors, the following equation must be satisfied according to (C2):
\begin{align}
    &\hat{\Pi}_{\mathcal{C}}\hat{Z}^{(j)}\hat{\Pi}_{\mathcal{C}}=\lambda_{j}\hat{\Pi}_{\mathcal{C}}~ \forall j \label{method:qecond12}
\end{align}
where $\lambda_{j}$'s are complex numbers. We note that Eq. \eqref{method:qecond12} can be directly derived from (C2) by setting $\hat{E}^{\dagger}_{i}=\hat{I} $ and $\hat{E}_{j}=\hat{Z}^{(j)} $. However, if both (C1) and (C2) (which implies Eq. \eqref{method:qecond12},) are satisfied, (C3) cannot be satisfied because
\begin{equation}
    \begin{split}
    &\langle \Psi_{\mathrm{L}} \vert \hat{H}^{2} \vert \Psi_{\mathrm{L}} \rangle-\langle \Psi_{\mathrm{L}} \vert \hat{H} \vert \Psi_{\mathrm{L}} \rangle^{2}\\
    &= \langle \Psi_{\mathrm{L}} \vert \left(\sum_{j=1}^{N}\lambda_{j}\hat{\Pi}_{\mathcal{C}}\right)^{2} \vert \Psi_{\mathrm{L}} \rangle-\langle \Psi_{\mathrm{L}} \vert \left(\sum_{j=1}^{N}\lambda_{j}\hat{\Pi}_{\mathcal{C}}\right) \vert \Psi_{\mathrm{L}} \rangle^{2}=0 \label{method:qfizero}
    \end{split}
\end{equation}
for all $\ket{\Psi_{\mathrm{L}}} \in \mathcal{C}$. Eq. \eqref{method:qfizero} can be easily derived using the following relations: $\hat{\Pi}_{\mathcal{C}}\ket{\Psi_{\mathrm{L}}}=\ket{\Psi_{\mathrm{L}}}$, $\hat{H}_{\mathrm{L}}=\sum_{i=1}^{N}\hat{Z}^{(i)}$, (C1) which is $[\hat{\Pi}_{\mathcal{C}},\hat{H}]=0$, and Eq. \eqref{method:qecond12}. Therefore, to satisfy (C3) at least one single Pauli $Z$ operator must not satisfy (C2). In other words, if a QEC code is capable of correcting all single Pauli $Z$ errors, it will also erase all the signal information. Conversely, if the code is designed to preserve signal information, it cannot correct the corresponding single Pauli $Z$ errors. In such a scenario, a single Pauli $Z$ error induces a logical error in the code space which cannot be corrected by QEC. (See Fig. \ref{fig:errorclass}.) Similarly, one can easily show that for the IIDP noise with $p_{x},p_{y} \neq 0$, no QEC scheme can correct all the single Pauli $X$ and $Y$ errors while simultaneously satisfying (C3).

\section{Author contributions}
H.K. conceived the main idea of combining virtual purification and quantum error correction in quantum metrology, performed the theoretical analysis and numerical simulations, prepared all figures, and wrote the manuscript.
C.O., Y.L., and H.J. contributed to the analytical derivations, assisted in developing the theoretical framework of virtual purification for metrological estimation, and proposed potential applications.
S.-W.L. supervised the project, provided critical insights into quantum error correction and quantum metrology, and revised the manuscript.
L.J. jointly supervised the project, contributed to theoretical discussions, and provided overall guidance and feedback on the manuscript.

\section{Acknowledgments}We acknowledge useful discussions with Seok-Hyung Lee, Seongjoo Cho, and Kento Tsubouchi. 

This research was funded by National Research Foundation of Korea (RS-2022-NR068812).
H.K. was supported by the IITP (RS-2025-02263264, RS-2025-25464252, RS-2024-00437191), the Education and Training Program of the Quantum Information Research Support Center (2021M3H3A1036573), and the NRF (RS-2025-25464492, RS-2024-00442710) funded by the Ministry of Science and ICT (MSIT), Korea.
H.J. was supported by the National Research Foundation of Korea (NRF) grant funded by the Korea government (MSIT) (RS-2024-00413957, RS-2024-00438415, RS-2023-NR076733), the Institute of Information \& Communications Technology Planning \& Evaluation (IITP) grant funded by the Korea government (MSIT) (IITP-2026-RS-2020-II201606, IITP-2026-RS-2024-00437191, and RS-2025-02219034), and the Institute of Applied Physics at Seoul National University. 
L.J. acknowledges support from the ARO(W911NF-23-1-0077), ARO MURI (W911NF-21-1-0325), AFOSR MURI (FA9550-19-1-0399, FA9550-21-1-0209, FA9550-23-1-0338), DARPA (HR0011-24-9-0359, HR0011-24-9-0361), NSF (OMA-1936118, ERC-1941583, OMA-2137642, OSI-2326767, CCF-2312755), NTT Research, and the Packard Foundation (2020-71479). 
Y.L. and C.O. acknowledge support by Quantum Technology R\&D Leading Program~(Quantum Computing) (RS-2024-00431768) through the National Research Foundation of Korea~(NRF) funded by the Korean government (Ministry of Science and ICT~(MSIT)) and the Institute of Information \& Communications Technology Planning \& Evaluation (IITP) Grants funded by the Korea government (MSIT) (No. IITP-2025-RS-2025-02283189). 
Y.L. acknowledges Institute of Information \& Communications Technology Planning \& Evaluation (IITP) grant funded by the Korea government (MSIT) (RS-2024-00437284, No. 2022-0-00463) and National Research Foundation of Korea(2023M3K5A109480511, RS-2023-NR119931).
C.O. was supported by the National Research Foundation of Korea Grants (No. RS-2025-00515456) funded by the Korean government (Ministry of Science and ICT (MSIT)) and the Institute of Information \& Communications Technology Planning \& Evaluation (IITP) Grants funded by the Korea government (MSIT) (No. IITP-2025-RS-2025-02263264 and IITP-2025-RS-2025-25464990).

\bibliography{Reference.bib}

\let\oldaddcontentsline\addcontentsline
\renewcommand{\addcontentsline}[3]{}
\let\addcontentsline\oldaddcontentsline
\onecolumngrid

\clearpage
\begin{center}
	\Large
	\textbf{Supplemental Material: Virtual purification complements quantum error correction in quantum metrology}
\end{center}

\setcounter{equation}{0}
\setcounter{figure}{0}
\setcounter{table}{0}

\setcounter{page}{1}
\renewcommand{\thesection}{S\arabic{section}}
\renewcommand{\theequation}{S\arabic{equation}}
\renewcommand{\thefigure}{S\arabic{figure}}
\renewcommand{\thetable}{S\arabic{table}}

\addtocontents{toc}{\protect\setcounter{tocdepth}{0}}

\setcounter{section}{0}

\section{Maximum Likelihood Estimator and Bias} \label{Appen A}
\subsection{Maximum Likelihood Estimator}
We introduce the \textit{maximum likelihood estimator} (MLE), a statistical method used to estimate parameters based on measurement outcomes. The conditional joint probability that we obtain measurement outcomes $\vec{A}=(A_{1},A_{2},\cdots , A_{M})$ when the signal parameter is a given value $\phi$, denoted as $P(\vec{A}\vert\phi)$, is called the \textit{likelihood function}. Here, $A_{i}$ is the $i$th measurement outcome. When each measurement outcome follows an independent identical distribution, the likelihood function is given by
\begin{align}
    P(\vec{A}\vert\phi) =\prod_{i=1}^{M}P(A_{i}\vert \phi).
\end{align}
The \textit{maximum likelihood estimation} adopts the value that maximizes the likelihood function as an estimator of $\phi$, i.e., 
\begin{align}
    \phi^{\text{est}}=\arg \max_{\theta} P(\vec{A}\vert\theta).
\end{align}
In the large sample number limit $M\to \infty$, the MLE exhibits two important properties. First, it is an unbiased estimator, $\left<\phi^{\text{est}}\right>=\phi$. Second, $\phi^{\text{est}}$ achieves the minimum statistical error (minimum variance), equal to the inverse of the Fisher information:
\begin{align}
    \V\left[\phi^{\text{est}}\right] = \frac{1}{F(\hat{A},\phi)},
\end{align}
where $F(\hat{A},\phi)$ is the Fisher information defined as
\begin{align}
    F(\hat{A},\phi)= \V\left[\pdv{}{\phi}\ln{P(\vec{A}\vert \phi)}\right]=\V\left[\pdv{}{\phi}\ln{\prod_{i=1}^{M}P(A_{i}\vert \phi)}\right]=\sum_{i=1}^{M}\V\left[\pdv{}{\phi}\ln{P(A_{i}\vert \phi)}\right]=M\V\left[\pdv{}{\phi}\ln{P(A\vert \phi)}\right]. \label{deffi}
\end{align}
We will show this below.

Next, let us inspect the MLE when the eigenvalues of an observable $\hat{A}$ are $1$ and $-1$ which is the case that we mainly focus on in this paper. We assume that the measurement outcome $A_{i}$ follows independent identical binomial distribution for all $i=1,2,\cdots,M$. In this case, since all the measurement outcomes follow the binomial distribution, 
\begin{align}
    P(\vec{A}\vert \theta)=\prod_{i=1}^{M}P(A_{i}\vert \theta)=p(\phi)^{n_{1}}(1-p(\phi))^{M-n_{1}},
\end{align}
where $p(\phi) \equiv \frac{1}{2}\left(1+\Tr\left[\hat{A}\hat{\rho}(\phi)\right]\right)$ is the probability that the measurement outcome is $1$, and $n_{1} \equiv \sum_{k=1}^{M}\left(\frac{A_{k}+1}{2}\right)$ is the number of $1$ among $M$ number of measurement outcomes. To find the MLE, let us find $\theta$ that maximizes $P(\vec{A}\vert \theta)$ using the \textit{log-likelihood function} which we denote as $L(\theta)\equiv \ln{P(\vec{A}\vert \theta)}$. At the parameter value that maximizes the likelihood function (equivalently, the value maximizes the log-likelihood function), the derivative of $L(\theta)$ becomes $0$:
\begin{align}
    \pdv{}{\theta}L(\theta)\bigg\vert_{\theta=\phi^{\text{est}}}=\pdv{p(\theta)}{\theta}\left[\frac{n_{1}-Mp(\theta)}{p(\theta)(1-p(\theta))}\right] \bigg\vert_{\theta=\phi^{\text{est}}}=0.
\end{align}
Therefore, when $\pdv{p(\theta)}{\theta}\neq 0$, $\phi^{\text{est}}$ satisfies
\begin{align}
    p(\phi^{\text{est}})=\frac{1}{2}\left(1+\mu(\phi^{\text{est}})\right)=\frac{n_{1}}{M}=\frac{1}{2}\left(1+\bar{A}\right),
\end{align}
which results in 
\begin{align}
    \phi^{\text{est}}=\mu^{-1}(\bar{A}), \label{mle22}
\end{align}
where $\mu(\phi)\equiv \Tr\left[\hat{A}\hat{\rho}(\phi)\right]$, $\bar{A}\equiv \sum_{i=1}^{M} A_{i}/M$, and $\mu^{-1}$ is the inverse function of $\mu(\phi)$. To find the asymptotic behavior of $\phi^{\text{est}}$ in the large $M$ limit, let us consider its Taylor's exapnsion
\begin{align}
    \phi^{\text{est}} =\mu^{-1}(\mu(\phi))+\sum_{k=1}^{\infty}M_{k}(\bar{A}-\mu(\phi))^{k}, \label{estimatorofphi}
\end{align}
where $M_{k}\equiv \frac{1}{k!}\pdv{^{k}}{x^{k}}\mu^{-1}(x)\big\vert_{x=\mu(\phi)}$. In the large sample limit, $\bar{A}$ follows the normal distribution $\mathcal{N}(\mu,\sigma)$ (because of the central limit theorem,) where $\mu(\phi) = \left<\bar{A}\right>=\Tr\left[\hat{A}\rho(\phi)\right]$ and $\sigma(\phi)^{2} \equiv \left<\bar{A}^{2}\right>-\left<\bar{A}\right>^{2}=\frac{1-\Tr\left[\hat{A}\rho(\phi)\right]^{2}}{M}$. According to the formulas of central moments of normal distributions \cite{supp-edition2002probability} which are
\begin{align}
    &\left<(\bar{A}-\mu)\right>^{2n}=\frac{(2n)!\sigma^{2n}}{n!2^{n}}, ~~\left<(\bar{A}-\mu)\right>^{2n+1}=0, \label{oddapprox}
\end{align}
the expectation value of the estimator is
\begin{align}
    \left<\phi^{\text{est}}\right> &= \mu^{-1}(\mu(\phi))+\sum_{k=1}^{\infty}f_{2k}\left<(\bar{A}-\mu(\phi))^{2k}\right>=\phi+\sum_{k=1}^{\infty}f_{2k}\frac{(2k)!}{k!2^{k}}\sigma(\phi)^{2k}. \label{expecMLE}
\end{align}
Lastly since $\sigma \to 0$ in the large sample limit $M \to \infty$,
\begin{align}
    \lim_{M\to \infty}\left<\phi^{\text{est}}\right>=\phi.
\end{align}

Using Eqs. \eqref{estimatorofphi} and \eqref{oddapprox}, one can find that the corresponding statistical error is
\begin{align}
    &\left<\left(\phi^{\text{est}}\right)^{2}\right>-\left<\left(\phi^{\text{est}}\right)\right>^{2}= \left( \pdv{}{x}\mu^{-1}(x)\big\vert_{x=\mu(\phi)}\right)^{2}\sigma^{2}= \left(\frac{1}{\pdv{\mu(\phi)}{\phi}}\right)^{2}\sigma(\phi)^{2}. \label{estvari}
\end{align}
Lastly, let us inspect whether the variance defined in Eq. \eqref{estvari} achieves the minimum statistical error which is the inverse of the Fisher information. When the eigenvalues of an observable $\hat{A}$ are $\pm 1$, the Fisher information defined in Eq. \eqref{deffi} is
\begin{align}
    &M\V\left[\pdv{}{\phi}\ln{P(A \vert \phi)}\right]=M\sum_{A}\frac{1}{P(A \vert \phi)}\left(\pdv{}{\phi}P(A \vert \phi)\right)^{2}-M\left[\sum_{A}\pdv{}{\phi}P(A \vert \phi)\right]^{2} = M\sum_{A}\frac{1}{P(A \vert \phi)}\left(\pdv{}{\phi}P(A \vert \phi)\right)^{2} \\
    &=M\left(\frac{1}{\frac{1+\mu(\phi)}{2}}\left[\pdv{}{\phi}\left(\frac{1+\mu(\phi)}{2}\right)\right]^{2}+\frac{1}{\frac{1-\mu(\phi)}{2}}\left[\pdv{}{\phi}\left(\frac{1-\mu(\phi)}{2}\right)\right]^{2}\right)=M\left(\pdv{\mu(\phi)}{\phi}\right)^{2}\frac{1}{\sigma(\phi)^{2}},
\end{align}
which is the inverse of the variance of the MLE.

\subsection{Error case}
Let us denote the noise-free quantum probe for $\phi$ estimation as $\ket{\psi_{0}}$. The quantum probe embeds $\phi$ where the signal state will be denoted as $
\dyad{\psi(\phi)}$. Here we call the noise-free signal state as ideal state. However, in practice, there are noises during the estimation. Furthermore, characterizing and calibrating the effect of noise is not always feasible in practice, which results in a bias of an estimator of $\phi$. 

The noisy quantum state in the $N$-qubit system can be described as
\begin{align}
    \mathcal{E}_{\Delta}(\dyad{\psi(\phi)})=\hat{\rho}_{\text{e}}\equiv \lambda \dyad{\psi_{\text{e}}}+(1-\lambda)\hat{\sigma}_{\text{e}},
\end{align}
where $\mathcal{E}_{\Delta}(\cdot)$ is a noise channel with noise strength $\Delta$, and $\lambda$ and $\ket{\psi_{\text{e}}}$ are the largest eigenvalue and the corresponding eigenvector satisfying $\lim_{\Delta \to 0}\lambda=1$ and $\lim_{\Delta \to 0}\ket{\psi_{\text{e}}}=\ket{\psi}$. $\hat{\sigma}_{\text{e}}$ is the density operator orthogonal to $\ket{\psi_{\text{e}}}$.
Due to the noise, measurement outcomes $\vec{A}_{\text{e}}\equiv(A_{\text{e},1},A_{\text{e},2},\cdots,A_{\text{e},M})$ and their average $\bar{A}_{\text{e}}=\sum_{k=1}^{M}A_{\text{e},k}/M$ are obtained instead of $\vec{A}$ and $\bar{A}$, where $A_{\text{e},i}$ is $i$th measurement outcome from $\hat{\rho}_{\text{e}}$. When the error state (except $\phi$) is fully characterized, an unbiased estimator analogous to Eq. \eqref{mle22} can be set. However, when the noise characterization is intractable, a practical choice of an estimator would be 
\begin{align}
    \phi^{\text{est}}_{\text{e}}\equiv \mu^{-1}(\bar{A}_{\text{e}}),     \label{esterror}
\end{align}
since $\mu(\phi)$ is known. In analogous to Eq. \eqref{expecMLE}, one can find that the expectation value of such an estimator in the large sample limit is 
\begin{align}
    \left<\phi^{\text{est}}_{\text{e}}\right>=\mu^{-1}(\mu_{\text{e}}(\phi)),
\end{align}
where $\mu_{\text{e}}\equiv \Tr[\hat{A}\hat{\rho}_{\text{e}}]$.
Therefore the bias of $\phi^{\text{est}}_{\text{e}}$ is
\begin{align}
    B_{\text{e}}&\equiv \left<\phi^{\text{est}}_{\text{e}}\right>-\phi=\mu^{-1}(\mu_{\text{e}}(\phi))-\mu^{-1}(\mu(\phi))=\sum_{k=1}^{\infty} M_{k}\left(\mu_{\text{e}}-\mu\right)^{k}=\frac{\mu_{\text{e}}-\mu}{\pdv{\mu}{\phi}}+O\left((\mu_{\text{e}}-\mu)^{2}\right).
\end{align}
where $M_{k}\equiv \frac{1}{k!}\pdv{^{k}}{x^{k}}\mu^{-1}(x)\big\vert_{x=\mu}$. Next, let us express the bias $B_{\text{e}}$ in terms of the noise strength $\Delta$. First, let us inspect $\mu_{\text{e}}(\phi)-\mu(\phi)$:
\begin{align}
    \begin{split}
    &\mu_{\text{e}}(\phi)-\mu(\phi)=\lambda \left[\langle \psi_{\text{e}}  \vert \hat{A} \vert \psi_{\text{e}} \rangle - \langle \psi \vert \hat{A} \vert \psi \rangle \right]+(1-\lambda)\left[\Tr[\hat{A}\hat{\sigma}_{\text{e}}] - \langle \psi \vert \hat{A} \vert \psi \rangle\right]\\
    &=\left[1 - \sum_{k=1}^{\infty}\lambda_{k}\Delta^{k}\right]\left[\sum_{k=1}^{\infty}a_{k}\Delta^{k}\right]+\left[\sum_{k=1}^{\infty}\lambda_{k}\Delta^{k}\right]\left[\sum_{k=0}^{\infty}b_{k}\Delta^{k}\right]=\sum_{k=1}^{\infty}\left[a_{k} + \sum_{l=1}^{k}\lambda_{l}a_{k-l} - \sum_{l=1}^{k}\lambda_{l}b_{k-l}\right]\Delta^{k}= \sum_{k=1}^{\infty}f_{k}\Delta^{k}, \label{mueorder}
    \end{split}
\end{align}
where the relevant terms $\lambda_{k}$, $a_{k}$, $b_{k}$ and, $f_{k}$ are defined as
\begin{align}
    &1-\lambda = \sum_{k=0}^{\infty}\lambda_{k}(\phi)\Delta^{k}, \label{lambdaterm}\\
    &\langle \psi_{\text{e}} \vert \hat{A} \vert \psi_{\text{e}} \rangle - \langle \psi \vert \hat{A} \vert \psi \rangle = \sum_{k=0}^{\infty}a_{k}(\phi)\Delta^{k}, \label{dominant}\\ 
    &\Tr[\hat{A}\hat{\sigma}_{\text{e}}] - \langle \psi \vert \hat{A} \vert \psi \rangle = \sum_{k=0}^{\infty}b_{k}(\phi)\Delta^{k}, \label{tailterm}\\
    &f_{k}(\phi)\equiv a_{k}(\phi) - \sum_{l=0}^{k}\lambda_{l}(\phi)a_{k-l}(\phi) + \sum_{l=0}^{k}\lambda_{l}(\phi)b_{k-l}(\phi).
\end{align}
Finally, the bias can be expressed as
\begin{align}
    B_{\text{e}}=\frac{1}{\frac{\mu(\phi)}{d\phi}}f_{1}\Delta+O(\Delta^{2}). \label{biaserror}
\end{align}

Next, let us find the statistical error $\left<\left(\phi^{\text{est}}_{\text{e}}\right)^{2}\right>-\left<\phi^{\text{est}}_{\text{e}}\right>^{2}$. Since $\phi^{\text{est}}_{\text{e}}=\mu^{-1}(\bar{A}_{\text{e}})$, can be expressed as
\begin{align}
    \phi^{\text{est}}_{\text{e}}=\mu^{-1}(\bar{A}_{\text{e}})=\sum_{k=0}^{\infty}M_{\text{e},k}(\bar{A}_{\text{e}}-\mu_{\text{e}}(\phi))^{k}.
\end{align}
Here $M_{\text{e},k}\equiv \frac{1}{k!}\pdv{^{k}}{x^{k}}\mu^{-1}(x)\big\vert_{x=\mu_{\text{e}}}$. Therefore, according to Eq. \eqref{oddapprox},
\begin{align}
    &\left<\left(\phi^{\text{est}}_{\text{e}}\right)^{2}\right>=\left(M_{\text{e},0}\right)^{2}+\left(2M_{\text{e},0}M_{\text{e},2}+M_{\text{e},1}M_{\text{e},1}\right)\sigma_{\text{e}}(\phi)^{2}+O\left(\frac{1}{\left(M\right)^{4}}\right),\\
    &\left<\phi^{\text{est}}_{\text{e}}\right>^{2}=\left(M_{\text{e},0}\right)^{2}+\left(2M_{\text{e},0}M_{\text{e},2}\right)\sigma_{\text{e}}(\phi)^{2}+O\left(\frac{1}{\left(M\right)^{4}}\right),
\end{align}
where $\sigma_{\text{e}}(\phi)^{2}\equiv \frac{\left<\bar{A}^{2}_{\text{e}}\right>-\left<\bar{A}_{\text{e}}\right>^{2}}{M} =\frac{1-\Tr\left[\hat{A}\hat{\rho}_{\text{e}}(\phi)\right]^{2}}{M}$. As a result, in the large sample limit $M \to \infty$, the statistical error is 
\begin{align}
    \left<\left(\phi^{\text{est}}_{\text{e}}\right)^{2}\right>-\left<\phi^{\text{est}}_{\text{e}}\right>^{2}=\left(\frac{1}{\pdv{\mu(\phi)}{\phi}\vert_{\phi=\phi_{\text{e}}}}\right)^{2}\sigma_{\text{e}}(\phi)^{2}, \label{staterrore}
\end{align}
where $\phi_{\text{e}}$ satisfies $\mu(\phi_{\text{e}})=\mu_{\text{e}}(\phi)$.

\subsection{VP case}
One can find that the average of $\bar{A}_{\text{mit}}$ over all possible measurement outcomes is $\left<\bar{A}_{\text{mit}}\right>=\Tr\left[\hat{A}\hat{\rho}_{\text{mit}}\right] \equiv \mu_{\text{mit}}$ where we name $\hat{\rho}_{\text{mit}}$ as mitigated state which is defined as
\begin{align}
    \hat{\rho}_{\text{mit}}&=\frac{\hat{\rho}_{\text{e}}^{n}}{\Tr\left[\hat{\rho}_{\text{e}}^{n}\right]}
    \equiv \frac{\lambda^{n}\dyad{\psi_{1}}+(1-\lambda)^{n}(\hat{\sigma}_{\text{e}})^{n}}{\lambda^{n}+(1-\lambda)^{n}\mathrm{Tr}[(\hat{\sigma}_{\text{e}})^{n}]}.
\end{align}
Similar to the error case, one can consider the following estimator for the mitigation case:
\begin{align}
    \phi^{\text{est}}_{\text{mit}}\equiv \mu^{-1}(\bar{A}_{\text{mit}}), \label{estmit}
\end{align}
whose expectation value is
\begin{align}
    \left<\phi^{\text{est}}_{\text{mit}}\right> = \mu^{-1}(\mu_{\text{mit}}(\phi)).
\end{align}
The bias of the corresponding estimator is
\begin{align}
    B_{\text{mit}}&\equiv \left<\phi^{\text{est}}_{\text{mit}}\right>-\phi=\mu^{-1}(\mu_{\text{mit}}(\phi))-\mu^{-1}(\mu(\phi)) = \sum_{k=1}^{\infty} M_{k}\left(\mu_{\text{mit}}-\mu\right)^{k}=\frac{\mu_{\text{mit}}-\mu}{\pdv{\mu}{\phi}}+O(\mu_{\text{mit}}-\mu)^{2},
\end{align}
where $M_{k}\equiv \frac{1}{k!}\pdv{^{k}}{x^{k}}\mu^{-1}(x)\big\vert_{x=\mu}$.
Next, let us inspect $\mu_{\text{mit}}(\phi)-\mu(\phi)$. For the simpler expression, let us define the coefficient $\lambda'$ whose Taylor expansion is
\begin{align}
    \begin{split}
    &\lambda'=\frac{\lambda^{n}}{\lambda^{n}+(1-\lambda)^{n}\Tr[(\hat{\sigma}_{\text{e}})^{n}]}=\frac{1}{1+\left(\frac{1-\lambda}{\lambda}\right)^{n}\Tr[(\hat{\sigma}_{\text{e}})^{n}]}=1+\sum_{k=1}^{\infty}\left[-\left(\frac{1-\lambda}{\lambda}\right)^{n}\Tr[(\hat{\sigma}_{\text{e}})^{n}]\right]^{k}\\
    &=1+\sum_{k=1}^{\infty}\left[-\left(\sum_{l=1}^{\infty}\lambda_{l}\Delta^{l}\right)^{n}\left(1+\sum_{l=1}^{\infty}\left[\sum_{l'=1}^{\infty}\lambda_{l'}\Delta^{l'}\right]^{l}\right)^{n}\Tr[(\hat{\sigma}_{\text{e}})^{n}]\right]^{k} \equiv 1-\sum_{k=n}^{\infty}\lambda'_{k}\Delta^{k},
    \end{split}
\end{align}
The $\mu_{\text{mit}}(\phi)-\mu(\phi)$ is then
\begin{align}
    \begin{split}
    &\mu_{\text{mit}}(\phi)-\mu(\phi)=\lambda' \left[\langle \psi_{\text{e}}  \vert \hat{A} \vert \psi_{\text{e}} \rangle - \langle \psi \vert \hat{A} \vert \psi \rangle \right]+(1-\lambda')\left[\Tr[\hat{A}(\hat{\sigma}_{\text{e}})^{n}] - \langle \psi \vert \hat{A} \vert \psi \rangle\right]\\
    &=\left[1 - \sum_{k=n}^{\infty}\lambda'_{k}\Delta^{k}\right]\left[\sum_{k=1}^{\infty}a_{k}\Delta^{k}\right]+\sum_{k=n}^{\infty}\lambda'_{k}\left[\sum_{k=0}^{\infty}c_{k}\Delta^{k}\right]\\
    &=\sum_{k=1}^{n-1}a_{k}\Delta^{k}+ \sum_{k=n}^{\infty} \left[a_{k} + \sum_{l=n}^{k}\lambda'_{l}a_{k-l} - \sum_{l=n}^{k}\lambda'_{l}c_{k-l}\right] \Delta^{k}\equiv \sum_{k=1}^{n-1}a_{k}\Delta^{k} + \sum_{k=n}^{\infty} g_{k}\Delta^{k} = \sum_{k=1}^{n-1}a_{k}\Delta^{k} + O(\Delta^{n}) \label{mumitorder}
    \end{split}
\end{align}
where the relevant term $c_{k}$ and $g_{k}$ are
\begin{align}
    &\Tr[\hat{A}(\hat{\sigma}_{\text{e}})^{n}] - \langle \psi \vert \hat{A} \vert \psi \rangle = \sum_{k=0}^{\infty}c_{k}\Delta^{k}, \\
    &g_{k} = a_{k} + \sum_{l=n}^{k}\lambda'_{l}a_{k-l} - \sum_{l=n}^{k}\lambda'_{l}c_{k-l}.
\end{align}
Finally, the bias can be expressed as
\begin{align}
    B_{\text{mit}}=\frac{1}{\frac{d\mu(\phi)}{d\phi}}\sum_{k=1}^{n-1}a_{k}\Delta^{k}+O(\Delta^{n}). \label{supple:biasmit}
\end{align}

Next, let us inspect the statistical error $\left<\left(\phi^{\text{est}}_{\text{mit}}\right)^{2}\right>-\left<\phi^{\text{est}}_{\text{mit}}\right>^{2}$. Following the same techniques that we used for the derivation of Eq. \eqref{staterrore}, one can find that
\begin{align}
    \left<\left(\phi^{\text{est}}_{\text{mit}}\right)^{2}\right>-\left<\phi^{\text{est}}_{\text{mit}}\right>^{2}=\left(\frac{1}{\pdv{\mu(\phi)}{\phi}\vert_{\phi=\phi_{\text{mit}}}}\right)^{2}\sigma_{\text{mit}}(\phi)^{2}
\end{align}
where $\phi_{\text{mit}}$ satisfies $\mu(\phi_{\text{mit}})=\mu_{\text{mit}}(\phi)$ and $\sigma_{\text{mit}}^{2}$ is the variance of $\bar{A}_{\text{mit}}$, which can be found in Refs. \cite{noi-yamamoto2021error,vp-PhysRevX.11.031057,vp-PhysRevX.11.041036,qem-kwon2024efficacy}, is
\begin{align}
    \sigma_{\text{mit}}^{2}\equiv \frac{\left<\bar{A}^{2}_{\text{mit}}\right>-\left<\bar{A}_{\text{mit}}\right>^{2}}{M}=\frac{2n}{M}\bigg[\frac{1-\Tr\left[\hat{A}\hat{\rho}^{n}_{\text{e}}\right]^{2}}{\Tr\left[\hat{\rho}^{n}_{\text{e}}\right]^{2}} +\frac{\Tr\left[\hat{A}\hat{\rho}^{n}_{\text{e}}\right]^{2} (1-\Tr\left[\hat{\rho}^{n}_{\text{e}}\right]^{2})}{\Tr\left[\hat{\rho}^{n}_{\text{e}}\right]^{4}} \bigg].\label{staterrormit}
\end{align}
\section{Bias of error case}
\subsection{error state}
Let us find the explicit form of the error state. Under the IIDP noise with noise distributions $p_{x}=k_{x}\Delta+O(\Delta^{2}), p_{y}=k_{y}\Delta+O(\Delta^{2}), p_{z}=k_{z}\Delta+O(\Delta^{2}), p_{I}=1-(k_{x}+k_{y}+k_{z})\Delta + O(\Delta^{2})$, the error state is
\begin{align}
    \hat{\rho}_{\text{e}}&=(1-\Delta(2Nk_{x}+2Nk_{y}+2Nk_{z}))\dyad{\psi}+\Delta(2Nk_{x}+2Nk_{y}+2Nk_{z})\hat{\sigma_{\text{e}}} +O(\Delta^{2})
\end{align}
where
\begin{align}
    (2Nk_{x}+2Nk_{y}+2Nk_{z})\hat{\sigma_{\text{e}}}&=k_{x}\sum_{k=1}^{N}\hat{X}^{(k)}\dyad{\psi_{s}}\hat{X}^{(k)}+k_{y}\sum_{k=1}^{N}\hat{Y}^{(k)}\dyad{\psi_{s}}\hat{Y}^{(k)}+2k_{z}\sum_{k=1}^{N}\hat{Z}^{(k)}\dyad{\psi_{s}}\hat{Z}^{(k)} \label{afterencoding}\\ 
    &+k_{x}\sum_{k=1}^{N}\hat{U}(\phi)\hat{X}^{(k)}\dyad{\psi_{s0}}\hat{X}^{(k)}\hat{U}^{\dagger}(\phi)+k_{y}\sum_{k=1}^{N}\hat{U}(\phi)\hat{Y}^{(k)}\dyad{\psi_{s0}}\hat{Y}^{(k)}\hat{U}^{\dagger}(\phi) +O(\Delta^{2}). \label{beforeencoding}
\end{align}
We note that the first and second terms in the right side in Eq. \eqref{afterencoding} denote the local Pauli noises that occur after the signal embedding, the last term in Eq. \eqref{afterencoding} has a factor of 2, including both before and after noise channel, and the terms in Eq. \eqref{beforeencoding} denote the local Pauli noise that occur before the signal embedding. In the main text, we have shown that $\hat{X}^{(i)}\ket{\psi_{s}}$, $\hat{Y}^{(i)}\ket{\psi_{s}}$, $\hat{Z}^{(i)}\ket{\psi_{s}}$, $\hat{U}(\phi)\hat{X}^{(i)}\ket{\psi_{s0}}$, and $\hat{U}(\phi)\hat{Y}^{(i)}\ket{\psi_{s0}}$, which are the first order of $\Delta$ components of $\hat{\rho}_{\text{e}}$, are orthogonal to $\ket{\psi_{s}}$ regardless of $\phi$ for all $1 \leq i \leq N$. Therefore, the corresponding bias is
\begin{align}
    B_{\text{e}}=\frac{1}{\pdv{\mu(\phi)}{\phi}} 2N(k_{x}+k_{y}+k_{z}) \left(\Tr\left[\hat{A}\hat{\sigma}_{\text{e}}\right]- \mu(\phi)\right)\Delta + O(\Delta^{2}) \label{supple:biase}
\end{align}

\subsection{Bias for local depolarizing and dephasing noise}
Next, let us find the bias $B_{\text{e}}$ when the noise type is local depolarizing or dephasing noise. We note that local depolarizing noise and dephasing noise can be described with the noise distributions $p_{x}=p_{y}=p_{z}=\Delta/4$ and $p_{x}=p_{y}=0$, $p_{z}=\Delta/2$ respectively. Therefore, lets us consider the noise distribution $p_{x}=p_{y}=k\Delta$, $p_{z}=k_{z}\Delta$ where $k,k_{z}\geq 0$, and $k+k_{z} > 0$. This noise distribution can describe both the local depolarizing and dephasing noises. To find $B_{\text{e}}$, let us consider a measurement observable $\hat{A}$. Since we are considering a Pauli measurement, $\hat{A}$ can be expressed as
\begin{align}
    \hat{A}=\prod_{k=1}^{N}(i)^{\left\lfloor\frac{\sigma^{(k)}_{x}+\sigma^{(k)}_{z}}{2}\right \rfloor}\left(\hat{X}^{(k)}\right)^{\sigma^{(k)}_{x}}\left(\hat{Z}^{(k)}\right)^{\sigma^{(k)}_{z}}
\end{align}
where $\sigma^{(k)}_{x},\sigma^{(k)}_{z} \in \{0,1\}$ and $\lfloor \cdot \rfloor$ is the floor function. For a simpler expression, let us define $\sigma^{(k)}_{y} \equiv \sigma^{(k)}_{x}+\sigma^{(k)}_{z}$. Here we note that if $\hat{A}$ commutes with $\hat{H}$, $\langle \psi_{s} \vert \hat{A} \vert \psi_{s} \rangle = 1$ regardless of the value of $\phi$. In this case, one cannot estimate $\phi$. Therefore, we consider $\hat{A}$ that does not commute with $\hat{H}$. This condition is equivalent to that $\hat{A}$ anti-commutes at least one of single Pauli $Z$ operators, which implies that $\sum_{k=1}^{N}\sigma^{(k)}_{x} > 0$. We assume such $\hat{A}$. We then find the relevant terms that contribute to $\Tr\left[\hat{A}\hat{\sigma}_{\text{e}}\right]$:
\begin{align}
    &\langle \psi_{s} \vert  \hat{X}^{(k)} \hat{A} \hat{X}^{(k)} \vert \psi_{s} \rangle = (-1)^{\sigma^{(k)}_{z}}\mu(\phi), \label{afterX}\\
    &\langle \psi_{s} \vert  \hat{Y}^{(k)} \hat{A} \hat{Y}^{(k)} \vert \psi_{s} \rangle = (-1)^{\sigma^{(k)}_{y}}\mu(\phi),\label{afterY}\\
    &\langle \psi_{s} \vert  \hat{Z}^{(k)} \hat{A} \hat{Z}^{(k)} \vert \psi_{s} \rangle = (-1)^{\sigma^{(k)}_{x}}\mu(\phi),\label{afterZ}\\
    &\langle \psi_{s0} \vert  \hat{X}^{(k)} \hat{U}^{\dagger}(\phi) \hat{A} \hat{U}(\phi) \hat{X}^{(k)} \vert \psi_{s0} \rangle = (-1)^{\sigma^{(k)}_{z}}\cos^{2}{\phi}\mu(\phi)+(-1)^{\sigma^{(k)}_{y}}\sin^{2}{\phi}\mu(\phi)+\cos{\phi}\sin{\phi}\langle \psi_{s} \vert  \hat{X}^{(k)} \hat{A} \hat{Y}^{(k)}+\hat{Y}^{(k)} \hat{A} \hat{X}^{(k)} \vert \psi_{s} \rangle,\label{beforeX}\\
    &\langle \psi_{s0} \vert  \hat{Y}^{(k)} \hat{U}^{\dagger}(\phi) \hat{A} \hat{U}(\phi) \hat{Y}^{(k)} \vert \psi_{s0} \rangle = (-1)^{\sigma^{(k)}_{z}}\sin^{2}{\phi}\mu(\phi)+(-1)^{\sigma^{(k)}_{y}}\cos^{2}{\phi}\mu(\phi)-\cos{\phi}\sin{\phi}\langle \psi_{s} \vert  \hat{X}^{(k)} \hat{A} \hat{Y}^{(k)}+\hat{Y}^{(k)} \hat{A} \hat{X}^{(k)} \vert \psi_{s} \rangle.\label{beforeY}
\end{align}
Using Eqs. \eqref{supple:biase} and \eqref{afterX}-\eqref{beforeY}, we find that the corresponding bias is
\begin{align}
    B_{\text{e}}=\frac{\mu_{\text{e}}(\phi)-\mu(\phi)}{\pdv{\mu(\phi)}{\phi}} +O(\Delta^{2}) = -\Delta \left[2k\sum_{k=1}^{N}\left(1-(-1)^{\sigma^{(k)}_{z}}\right)+2k\sum_{k=1}^{N}\left(1-(-1)^{\sigma^{(k)}_{y}}\right)+2k_{z}\sum_{k=1}^{N}\left(1-(-1)^{\sigma^{(k)}_{x}}\right)\right]\frac{\mu(\phi)}{\pdv{\mu(\phi)}{\phi}} + O(\Delta^{2})
\end{align}

\subsection{error corrected state}
Let us consider stabilizer codes $\mathcal{C}(\mathbf{s})$ which is defined with a stabilizer generator set $\mathbf{s}$. To satisfy the error correction condition (1): $[\hat{\Pi}_{\mathcal{C}},\hat{H}]=0$ where $\hat{\Pi_{\mathcal{C}}}$ is the projection operator on the code space, all the stabilizers generated by $\mathbf{s}$ should commutes with $\hat{H}$. Equivalently, $\forall \hat{S} \in \left<\mathbf{s}\right>$ commutes with all the single Pauli $Z$ operators, i.e, $[\hat{Z}^{(i)},\hat{H}]=0$ for all $i$'s. In such code, one cannot distinguish $\hat{I}$ and $\hat{Z}^{(i)}$ for all $i$'s, i.e., all the single Pauli $Z$ errors give the same syndrome measurement with the noiseless case. As a consequence, such code cannot correct all the single Pauli $Z$ errors. Next, let us assume that the code distance of $\mathcal{C}(\mathbf{s})$ is larger than $1$. In such cases, the single Pauli $X$ errors take the $\mathcal{C}(\mathbf{s})$ to the orthogonal spaces, equivalently, such code can correct all the single Pauli $X$ errors. However, since $\hat{Y}^{(i)}=-i\hat{X}^{(i)}\hat{Z}^{(i)}$, one cannot distinguish $\hat{Y}^{(i)}$ and $\hat{X}^{(i)}$ through syndrome measurements. To get a better result, one can correct one of $\hat{X}^{(i)}$ and $\hat{Y}^{(i)}$ which has a higher probability. We note considering up to the first order of $\Delta$, this corresponds to the maximum likelihood decoder. that For example, when $p_{y}>p_{x}$, applying $\hat{Y}^{(i)}$ when one gets error syndrome that corresponds to $\hat{X}^{(i)}$. In such scenario, an error-corrected state is
\begin{align}
    \hat{\rho}_{\text{QEC}} \equiv \mathcal{R}\circ \mathcal{E}\circ\mathcal{U}_{\phi}(\dyad{\psi_{\text{L}0}})=(1-Np_{z}-N\min\{p_{x},p_{y}\})\dyad{\psi_{\text{L}}}+\left(p_{z}+\min\{p_{x},p_{y}\}\right)\sum_{i=1}^{N}\hat{Z}^{(i)}\dyad{\psi_{\text{L}}}\hat{Z}^{(i)}+O(\Delta^{2}),
\end{align}
and the corresponding bias is
\begin{align}
    B_{\text{QEC}}= -\Delta \left[\left(\min\{p_{x},p_{y}\}+p_{z}\right)\sum_{k=1}^{N}\left(1-(-1)^{\sigma^{(k)}_{x}}\right)\right]\frac{\mu(\phi)}{\pdv{\mu(\phi)}{\phi}} + O(\Delta^{2}).
\end{align}
Here, we emphasize that except the values of $\phi$ satisfying $\mu(\phi)=0$, $B_{\text{QEC}}=\Theta(\Delta)$.

\section{Higher order suppression}
In this section, we show that the bias of the VP can be further suppressed beyond $O(\Delta^{2})$ in the presence of the local dephasing error and depolarizing noise. We numerically studied such scenarios in Sec. \ref{numeric}.

\subsection{Local dephasing noise}\label{localdephasingsub}
Let us inspect the logical Pauli $Z$ operators acting on the code space $\mathcal{D}(\mathbf{s}^{[\text{c}]})$. We note that $\forall \hat{s}^{[\text{c}]} \in \mathbf{s}^{[\text{c}]}$ only consists of the products of the single Pauli $Z$ operators since all the CSs commutes with $\hat{H}$ and the number of generators is $(N-k)$. Therefore, there exists at least $k$-number of single Pauli $Z$ operators that cannot be generated by $\mathbf{s}^{[\text{c}]}$. Without loss of generality, let us assume that the corresponding single Pauli $Z$ operators are $\hat{Z}^{(1)},\hat{Z}^{(2)},\cdots \hat{Z}^{(k)}$. Since $\mathbf{s}^{[\text{c}]}\cup \{\hat{Z}^{(1)},\hat{Z}^{(2)},\cdots ,\hat{Z}^{(k)}\}$ forms an independent and commuting set, we consider $\hat{Z}^{(i)}$ as the logical single Pauli $Z$ acting on $i$th logical qubit. We denote the logical single Pauli $Z$ as $\hat{Z}^{(i)}_{\text{L}} \equiv \hat{Z}^{(i)}$ for all $i=1,2,\cdots ,k$. The set of the logical single Pauli $Z$ operators, which we denote as
\begin{align}
    \mathbf{z}_{\text{L}}=\{\hat{Z}^{(1)}_{\text{L}},\hat{Z}^{(2)}_{\text{L}},\cdots ,\hat{Z}^{(k)}_{\text{L}}\},
\end{align}
is the generator set of the logical Pauli $Z$ operators. We note that $\hat{Z}^{(i)}_{\text{L}}\ket{\psi_{s0}}$ is orthogonal to $\ket{\psi_{s0}}$ since $\hat{Z}^{(i)}_{\text{L}} \notin \left<\mathbf{s}\right>$ for all $1 \leq i \leq N$. Here $\big< \cdot \big>$ denotes a group that is generated by $(\cdot)$. As a consequence, $\forall \hat{Z} \in \left<\mathbf{s}^{[\text{c}]}\cup \mathbf{z}_{\text{L}}\right>$, $\hat{Z}\ket{\psi_{s0}}$ satisfies either $\hat{Z}\ket{\psi_{s0}}=\ket{\psi_{s0}}$ or $\langle \psi_{s0} \vert \hat{Z} \vert \psi_{s0} \rangle = 0$. We note that $\forall \hat{Z} \in \left<\mathbf{s}^{[\text{c}]}\cup \mathbf{z}_{\text{L}}\right>$, $\hat{Z}\ket{\psi_{s0}} \in \mathcal{D}(\mathbf{s}^{[\text{c}]})$ and because of the same reason, $\hat{Z}\hat{U}(\phi)\ket{\psi_{s0}} \in \mathcal{D}(\mathbf{s}^{[\text{c}]})$. As a consequence, $\hat{Z}\ket{\psi_{s}}=\hat{Z}\hat{U}(\phi)\ket{\psi_{s0}}=\hat{U}(\phi)\hat{Z}\ket{\psi_{s0}}$ also satisfies either $\hat{Z}\ket{\psi_{s}}=\ket{\psi_{s}}$ or $\langle \psi_{s} \vert \hat{Z} \vert \psi_{s} \rangle = 0$. 

Finally, let us consider a local dephasing error where the corresponding error state can be expressed as
\begin{align}
    \hat{\rho}_{\text{e}}(\phi)=p_{0}\dyad{\psi_{s}}+\sum_{k=1}^{2^{N}-1}p_{i}\hat{Z}_{i}\dyad{\psi_{s}}\hat{Z}_{i}, \label{dephasingerrostate}
\end{align}
Here, $\hat{Z}_{i}$'s are $N$-qubit Pauli $Z$ operators which are elements of $  \left<\mathbf{s}^{[\text{c}]}\cup \mathbf{z}_{\text{L}}\right>$ for all $i$'s.
As a consequence, for the local dephasing error, all the terms in Eq. \eqref{dephasingerrostate} are either $\ket{\psi_{s}}$ or orthogonal to $\ket{\psi_{s}}$, which implies the dominant eigenvector of $\hat{\rho}_{\text{e}}$ is $\ket{\psi_{s}}$. Therefore, $a_{k}$'s defined in Eq. \eqref{dominant} are $0$ for all $k$, which implies that $B_{\text{mit}}=O(\Delta^{n})$ where $n$ is the mitigation order of the VP. As a consequence, one can arbitrarily reduce bias when one has sufficient resources for the VP.

\subsection{Local depolarizing noise} \label{localdepolasub}
Let us consider the case when the code distance of $\mathcal{D}(\mathbf{s}^{[\text{c}]})$ is larger than $2$. We already have seen that the Pauli errors that corresponds to first order of $\Delta$, which are $\hat{X}^{(i)}\ket{\psi_{s}}$, $\hat{Y}^{(i)}\ket{\psi_{s}}$, $\hat{U}(\phi)\hat{X}^{(i)}\ket{\psi_{s0}}$, $\hat{U}(\phi)\hat{Y}^{(i)}\ket{\psi_{s0}}$, $\hat{Z}^{(i)}\ket{\psi_{s}}$ are orthogonal to $\ket{\psi_{s}}$. 

Next, let us inspect the Pauli errors that correspond to the second order of $\Delta$. One can easily show that when the local depolarizing noise occurs only after, where the corresponds error state can be written as  
\begin{align}
    \hat{P}^{(i)}\hat{Q}^{(j)} \ket{\psi_{s}}~~\text{or}~~ \hat{U}(\phi)\hat{P}^{(i)}\hat{Q}^{(j)} \ket{\psi_{s0}}, ~~~(i \neq j) \label{aftersignal}
\end{align}
where $\hat{P}^{(i)} \in \{\hat{X}^{(i)},\hat{Y}^{(i)},\hat{Z}^{(i)}\}$ and $\hat{Q}^{(j)} \in \{\hat{X}^{(j)},\hat{Y}^{(j)},\hat{Z}^{(j)}\}$, the corresponding error occurred states are either orthogonal to $\ket{\psi_{s}}$ or same with $\ket{\psi_{s}}$. The error occurred state can be same with $\ket{\psi_{s}}$ when $\hat{P}^{(i)}=\hat{Z}^{(i)}$ and $\hat{Q}^{(j)}=\hat{Z}^{(j)}$, and $\hat{Z}^{(i)}\hat{Z}^{(j)} \in \left<\mathbf{s}^{[\text{c}]}\right>$. 

Next, let us consider when the Pauli noise occurs on both sides of the signal embedding where  
corresponds error state can be written as  
\begin{align}
    \hat{P}^{(i)}\hat{U}(\phi)\hat{Q}^{(j)} \ket{\psi_{s0}}.
\end{align}
This case also, one can easily show that the corresponding error occurred states are either orthogonal to $\ket{\psi_{s}}$ or same with $\ket{\psi_{s}}$, except when $i=j$ and $\hat{P}^{(i)} \in \{\hat{X}^{(i)},\hat{Y}^{(i)}\}$ and $\hat{Q}^{(i)} \in \{\hat{X}^{(i)},\hat{Y}^{(i)}\}$. The corresponding error terms are
\begin{align}
    &\hat{X}^{(i)}\hat{U}(\phi)\hat{X}^{(i)}\ket{\psi_{s0}}= \hat{X}^{(i)}\left(\hat{X}^{(i)}\cos{\phi}+\hat{Y}^{(i)}\sin{\phi}\right)\ket{\psi_{s}}=\left(\hat{I}\cos{\phi}+i\hat{Z}^{(i)}\sin{\phi}\right)\ket{\psi_{s}}, \label{xxerror}\\
    &\hat{X}^{(i)}\hat{U}(\phi)\hat{Y}^{(i)}\ket{\psi_{s0}}= \hat{X}^{(i)}\left(-\hat{X}^{(i)}\sin{\phi}+\hat{Y}^{(i)}\cos{\phi}\right)\ket{\psi_{s}}=\left(-\hat{I}\sin{\phi}+i\hat{Z}^{(i)}\cos{\phi}\right)\ket{\psi_{s}},\\
    &\hat{Y}^{(i)}\hat{U}(\phi)\hat{X}^{(i)}\ket{\psi_{s0}}= \hat{Y}^{(i)}\left(\hat{X}^{(i)}\cos{\phi}+\hat{Y}^{(i)}\sin{\phi}\right)\ket{\psi_{s}}=\left(-i\hat{Z}^{(i)}\cos{\phi}+\hat{I}\sin{\phi}\right)\ket{\psi_{s}},\\
    &\hat{Y}^{(i)}\hat{U}(\phi)\hat{Y}^{(i)}\ket{\psi_{s0}}= \hat{Y}^{(i)}\left(-\hat{X}^{(i)}\sin{\phi}+\hat{Y}^{(i)}\cos{\phi}\right)\ket{\psi_{s}}=\left(i\hat{Z}^{(i)}\sin{\phi}+\hat{I}\sin{\phi}\right)\ket{\psi_{s}}, \label{YYerror}\\
\end{align}
which have probability $\left(\frac{\Delta}{4}\right)$
We note that each term in Eqs. \eqref{xxerror}-\eqref{YYerror} neither orthogonal to $\ket{\psi_{s}}$ nor same with $\ket{\psi_{s}}$.
However, sum of the terms can be represented as, 
\begin{align}
    &\mathfrak{R}(\hat{X}^{(i)}\hat{U}(\phi)\hat{X}^{(i)}\ket{\psi_{s0}})+\mathfrak{R}(\hat{X}^{(i)}\hat{U}(\phi)\hat{Y}^{(i)}\ket{\psi_{s0}})+\mathfrak{R}(\hat{Y}^{(i)}\hat{U}(\phi)\hat{X}^{(i)}\ket{\psi_{s0}})+\mathfrak{R}(\hat{Y}^{(i)}\hat{U}(\phi)\hat{X}^{(i)}\ket{\psi_{s0}})\\
    &= 2\dyad{\psi_{s}}+2\hat{Z}^{(i)}\dyad{\psi_{s}}\hat{Z}^{(i)}.
\end{align}
where we define following notation $\mathfrak{R}(\ket{\psi}) \equiv \dyad{\psi}$ for a simpler expression.

As a consequence, all the second-order errors terms which are $\hat{P}^{(i)}\hat{Q}^{(j \neq i)}\ket{\psi_{s}}$ $\hat{U}(\phi)\hat{P}^{(i)}\hat{Q}^{(j\neq i)}\ket{\psi_{s0}}$ $\hat{P}^{(i)}\hat{U}(\phi)\hat{Q}^{(j)}\ket{\psi_{s0}}$, either orthogonal to $\ket{\psi_{s}}$ or same with $\ket{\psi_{s}}$. This result implies that  $a_{2}$ defined in Eq. \eqref{dominant} are also $0$ where the $B_{\text{mit}}=O(\Delta^{3})$ when $n\geq 3$.

\section{Numerical simulations for stabilizer state quantum probes} \label{numeric}
For the numerical simulations, we investigate three distinct quantum probes: (1) a 5-qubit GHZ state, (2) a 5-qubit twin graph state having two pairs of true twins, and (3) the logical $0$ state of the $7$-qubit Steane code, all subjected to local depolarizing noise characterized by $p_{x} = p_{y} = p_{z} = \Delta/4$. For the quantum error correction (QEC) scenarios, the same quantum probes are utilized with the aid of an ancillary system. Specifically, we consider
\begin{align}
    \ket{\psi_{s0}}=\sum_{i=1}^{2^{N-k}}s_{i}\ket{i}_{\mathcal{D}}, \quad \ket{\psi_{\text{L}0}}=\sum_{i=1}^{2^{N-k}}s_i\ket{i}_{\mathcal{D}}\ket{i}_{\text{A}}
\end{align}
where $\{\ket{i}_{\mathcal{D}}\}_{i=1}^{2^{N-k}}$ and $\{\ket{i}_{\text{A}}\}_{i=1}^{2^{N-k}}$ denote the basis sets of the code space $\mathcal{D}(\mathbf{s}^{[\text{c}]})$ and the ancillary system, respectively, with the dimension of the code space given by $\text{dim}(\mathcal{D}(\mathbf{s}^{[\text{c}]}))=2^{N-k}$. We explore three noise strengths, corresponding to dominant eigenvalues of the error states at $0.8$, $0.7$, and $0.6$. The results for each quantum probe are illustrated in Figs. \ref{fig:suppleGHZPauli}, \ref{fig:TwinPauli}, and \ref{fig:SteanePauli}, corresponding to each probe. Across all simulations, we observe that the VP, with a mitigation order of $n=2,3$, efficiently reduces bias induced by the local depolarizing noise, whereas the QEC method can only reduce bias up to constant factors. These findings strongly support our main results. The detailed settings for each scenario are as follows:

\subsection{5-qubit GHZ state}
In the first scenario (in Fig. \ref{fig:suppleGHZPauli}), we consider $5$-qubit GHZ state where the CSGs and NCSGs are
\begin{align}
    \mathbf{s}^{[\text{c}]}=\{\hat{Z}^{(1)}\hat{Z}^{(2)},\hat{Z}^{(2)}\hat{Z}^{(3)},\hat{Z}^{(3)}\hat{Z}^{(4)},\hat{Z}^{(4)}\hat{Z}^{(5)}\},~~ \mathbf{s}^{[\text{n}]}=\{\hat{X}^{(1)}\hat{X}^{(2)}\hat{X}^{(3)}\hat{X}^{(4)}\hat{X}^{(5)}\}.
\end{align}
We consider $\hat{A} \equiv \prod_{i=1}^{5}\hat{Y}^{(i)}$ as a measurement observable that achieves the minimum statistical error in the absence of noise.

\begin{figure*}[t]
    \centering
    \includegraphics[width=\textwidth]{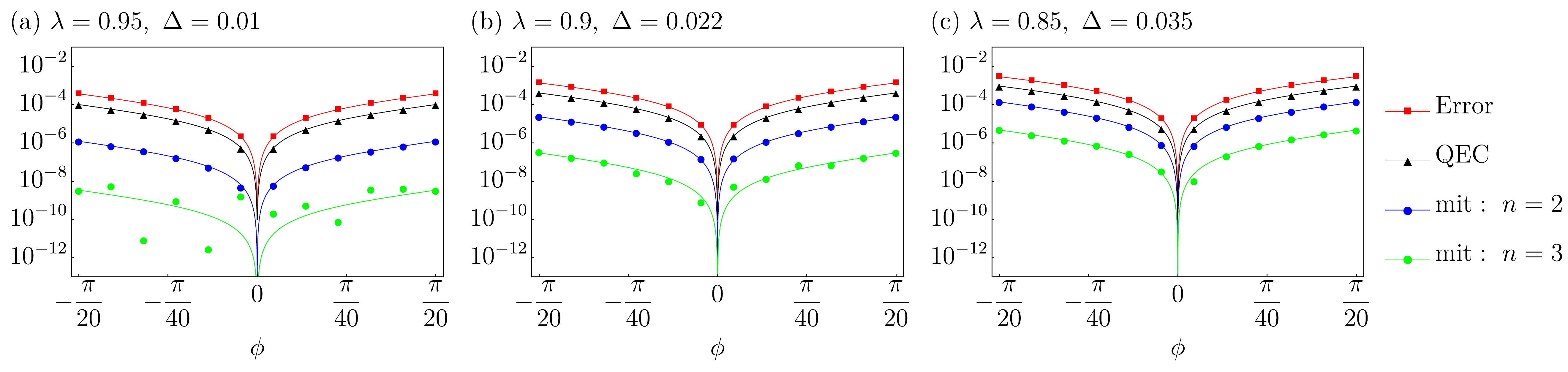}
    \caption{(a)-(c) 
     The square of biases (with a logarithmic scale) of the $\phi$ estimation, where $\phi \in \left[-\frac{\pi}{20}, \frac{\pi}{20}\right]$, exploiting $5$-qubit GHZ state as a quantum probe in the presence of local dephasing noise with different noise strengths (represented by different dominant eigenvalues). Solid lines indicate the theoretical values of the square of biases, while circles, squares, and up-triangles represent the squared differences between the estimated values and the actual values from simulations, i.e., $(\phi^{\text{est}} - \phi)^2$, where $\phi^{\text{est}}$ are MLE obtained with $M = 10^9$ samples. We emphasize that due to the large sample size, which significantly reduces fluctuations in $\phi^{\text{est}}$, the quantities $(\phi^{\text{est}} - \phi)^2$ can be interpreted as the square of bias derived from the simulations.
    } 
    \label{fig:GHZDephasing}
\end{figure*}

\begin{figure*}[t]
    \centering
    \includegraphics[width=\textwidth]{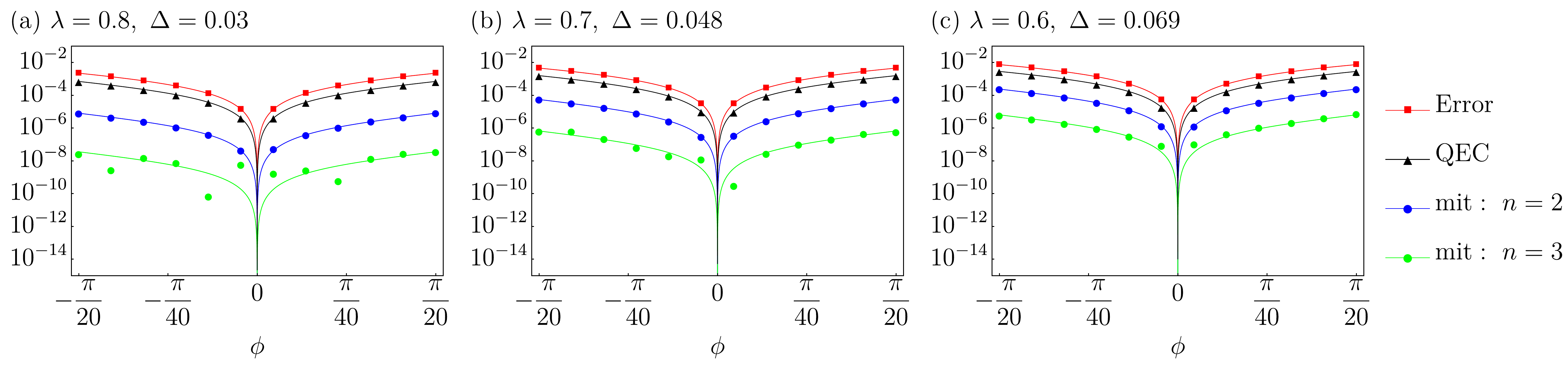}
    \caption{(a)-(c) 
     The square of biases (with a logarithmic scale) of the $\phi$ estimation, where $\phi \in \left[-\frac{\pi}{20}, \frac{\pi}{20}\right]$, exploiting $5$-qubit GHZ state in the presence of the local depolarizing noise. Other features are consistent with those in Fig. \ref{fig:GHZDephasing}.
    } 
    \label{fig:suppleGHZPauli}
\end{figure*}

\subsection{5-qubit Twin graph}
In this section, we introduce the twin graph state that we exploited in the numerical simulations. Twin graphs have been identified as quantum probes that facilitate quantum-enhanced estimation. We consider $5$-qubit twin graph which has two pairs of true twins: $2$nd and $5$th and $3$rd and $4$th qubits. (See Fig.) In a twin graph, each pair of true twins is connected to the same vertices.

\begin{figure}[t]
    \centering
    \includegraphics[width=0.3\linewidth]{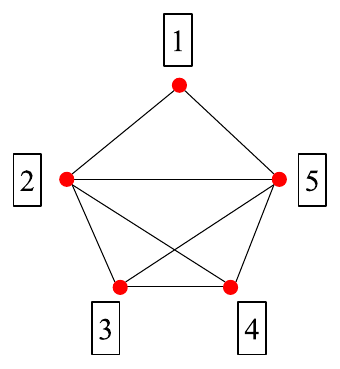}
    \caption{The graph that corresponds to $5$-qubit twin graph with the two pairs of true twins. Each number represents the qubit.} 
    \label{fig:twingraph}
\end{figure}

The corresponding stabilizer generator set for this state is given by
\begin{align}
    \mathbf{s}=\{\hat{X}^{(1)}\hat{Z}^{(2)}\hat{Z}^{(5)},\hat{Z}^{(1)}\hat{X}^{(2)}\hat{Z}^{(3)}\hat{Z}^{(4)}\hat{Z}^{(5)},\hat{Z}^{(2)}\hat{X}^{(3)}\hat{Z}^{(4)}\hat{Z}^{(5)},\hat{Z}^{(2)}\hat{Z}^{(3)}\hat{X}^{(4)}\hat{Z}^{(5)},\hat{Z}^{(1)}\hat{Z}^{(2)}\hat{Z}^{(3)}\hat{Z}^{(4)}\hat{X}^{(5)}\}.
\end{align}
However, the twin graph defined with $\mathbf{s}$ itself does not exhibit quantum-enhanced estimation precision. To address this, we align the quantum state by applying the Clifford operation $\hat{C} \equiv \prod_{i=1}^{5}e^{-i\frac{\pi}{4}\hat{X}^{(i)}}$. This transforms the stabilizer generator set into
\begin{align}
    \hat{C}\mathbf{s}\hat{C}^{\dagger}=\{\hat{X}^{(1)}\hat{Y}^{(2)}\hat{Y}^{(5)},\hat{Y}^{(1)}\hat{X}^{(2)}\hat{Y}^{(3)}\hat{Y}^{(4)}\hat{Y}^{(5)},\hat{Y}^{(2)}\hat{X}^{(3)}\hat{Y}^{(4)}\hat{Y}^{(5)},\hat{Y}^{(2)}\hat{Y}^{(3)}\hat{X}^{(4)}\hat{Y}^{(5)},\hat{Y}^{(1)}\hat{Y}^{(2)}\hat{Y}^{(3)}\hat{Y}^{(4)}\hat{X}^{(5)}\}. \label{sgtwin}
\end{align}
However, the stabilizer generators in Eq. \eqref{sgtwin} do not explicitly reveal the CSGs NCSGs. To uncover these, we consider an alternative set of stabilizer generators that generate the same group as $\hat{C}\mathbf{s}\hat{C}^{\dagger}$, 
\begin{align}
    \{\hat{Z}^{(2)}\hat{Z}^{(5)},\hat{Z}^{(3)}\hat{Z}^{(4)},\hat{X}^{(2)}\hat{X}^{(3)}\hat{Y}^{(4)}\hat{X}^{(5)}, \hat{X}^{(1)}\hat{Y}^{(2)}\hat{Y}^{(5)},\hat{Y}^{(1)}\hat{Y}^{(2)}\hat{Y}^{(3)}\hat{Y}^{(4)}\hat{X}^{(5)}\}. \label{csgncsgform}
\end{align}
We emphasize that the stabilizer generators in Eqs. \eqref{sgtwin} and \eqref{csgncsgform} generate the same stabilizer group. Finally, the CSGs and NCSGs are identified as 
\begin{align}
    \mathbf{s}^{[\text{c}]}=\{\hat{Z}^{(2)}\hat{Z}^{(5)},\hat{Z}^{(3)}\hat{Z}^{(4)}\},~~~\mathbf{s}^{[\text{n}]}=\{\hat{X}^{(2)}\hat{X}^{(3)}\hat{Y}^{(4)}\hat{X}^{(5)}, \hat{X}^{(1)}\hat{Y}^{(2)}\hat{Y}^{(5)},\hat{Y}^{(1)}\hat{Y}^{(2)}\hat{Y}^{(3)}\hat{Y}^{(4)}\hat{X}^{(5)}\}.
\end{align}

Next, we examine the quantum Fisher information (QFI) of the 5-qubit twin graph state and identify a measurement that achieves the QFI. The stabilizer state defined by the stabilizer generator set $\mathbf{s}^{[\text{c}]} \cup \mathbf{s}^{[\text{n}]}$ is
\begin{align}
    \ket{\psi_{0}} \equiv \frac{1}{2\sqrt{2}}\left(\ket{00000}-\ket{00110}+i\ket{01001}+i\ket{01111}-i\ket{10000}-i\ket{10110}-\ket{11001}+\ket{11111}\right).
\end{align}
The QFI of the canonical phase estimation using the twin graph state (without noise) can be calculated as
\begin{align}
    \langle \psi_{s0} \vert \hat{H}^{2} \vert \psi_{s0} \rangle -\langle \psi_{s0} \vert \hat{H} \vert \psi_{s0} \rangle^{2} = 9, \label{twinQFI}
\end{align}
where $\hat{H}=\sum_{i=1}^{N}\hat{Z}^{(i)}$.

To demonstrate that the measurement observable $\hat{A} \equiv \hat{Y}^{(1)}\hat{Y}^{(2)}\hat{Y}^{(3)}\hat{Y}^{(4)}\hat{X}^{(5)}$ achieves the QFI in Eq. \eqref{twinQFI}, in the small signal $\phi$ limit, we note that $\hat{A}$ anti-commutes with $\hat{H}$. Consequently, the expectation value of $\hat{A}$ over the signal state $\ket{\psi_{s}}$ and its derivative can be expressed in the small limit of $\phi$ as
\begin{align}
    &\bra{\psi_{s}} e^{i\hat{H}\frac{\phi}{2}} \hat{A} e^{-i\hat{H}\frac{\phi}{2}} \ket{\psi_{s}}= \bra{\psi_{s}} e^{i\hat{H}\phi} \ket{\psi_{s}} \approx 1+i\phi\bra{\psi_{s}}\hat{H} \ket{\psi_{s}}-\frac{1}{2}\phi^{2}\bra{\psi_{s}}\hat{H}^{2} \ket{\psi_{s}}= 1-\frac{1}{2}\phi^{2}\bra{\psi_{s}}\hat{H}^{2} \ket{\psi_{s}}, \label{expectationofA}\\
    &\pdv{}{\phi}\bra{\psi_{s}} e^{i\hat{H}\phi} \hat{A} e^{-i\hat{H}\phi} \ket{\psi_{s}}=-\phi \bra{\psi_{s}}\hat{H}^{2} \ket{\psi_{s}}.
\end{align}
Thus, according to the error propagation theorem, the ultimate statistical error with the measurement $\hat{A}$ is 
\begin{align}
     \sigma^{2}(\phi)=\frac{\bra{\psi_{s}} e^{i\hat{H}\frac{\phi}{2}} \hat{A}^{2} e^{-i\hat{H}\frac{\phi}{2}} \ket{\psi_{s}}-\bra{\psi_{s}} e^{i\hat{H}\frac{\phi}{2}} \hat{A} e^{-i\hat{H}\frac{\phi}{2}} \ket{\psi_{s}}^{2}}{\abs{\pdv{}{\phi}\bra{\psi_{s}} e^{i\hat{H}\frac{\phi}{2}} \hat{A} e^{-i\hat{H}\frac{\phi}{2}} \ket{\psi_{s}}}^{2}}=\frac{1}{\bra{\psi_{s}}\hat{H}^{2} \ket{\psi_{s}}}= \frac{1}{(\bra{\psi_{s}}\hat{H}^{2} \ket{\psi_{s}}-\bra{\psi_{s}}\hat{H} \ket{\psi_{s}}^{2})}.
\end{align}

\begin{figure*}[t]
    \centering
    \includegraphics[width=\textwidth]{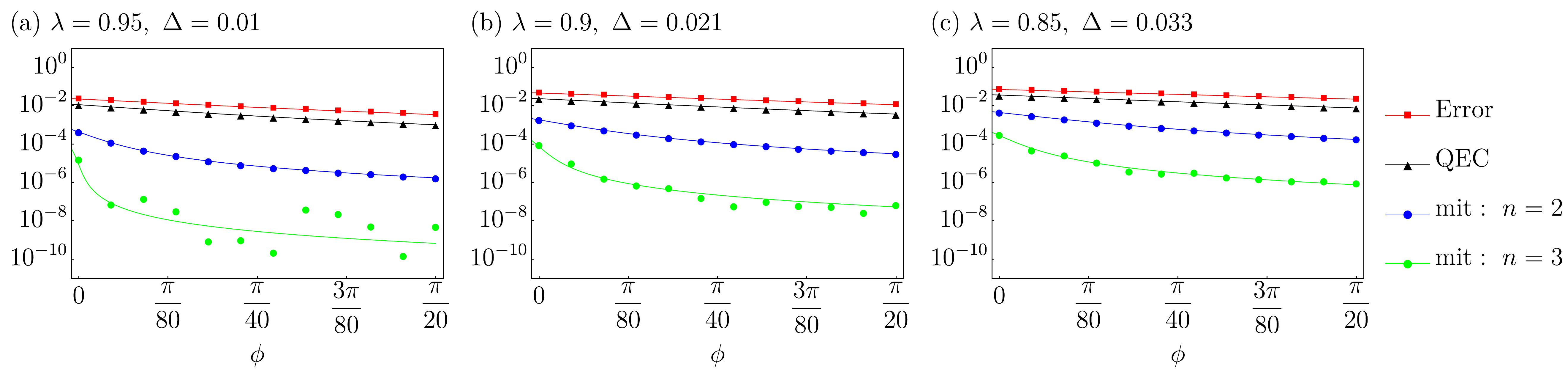}
    \caption{(a)-(c) 
    The square of biases (with a logarithmic scale) of the $\phi$ estimation, where $\phi \in \left[0, \frac{\pi}{20}\right]$, exploiting $5$-qubit Twin graph state with two pairs of true twins in the presence of the local dephasing noise. Other features are consistent with those in Fig. \ref{fig:GHZDephasing}.
    } 
    \label{fig:TwinDephasing}
\end{figure*}

\begin{figure*}[t]
    \centering
    \includegraphics[width=\textwidth]{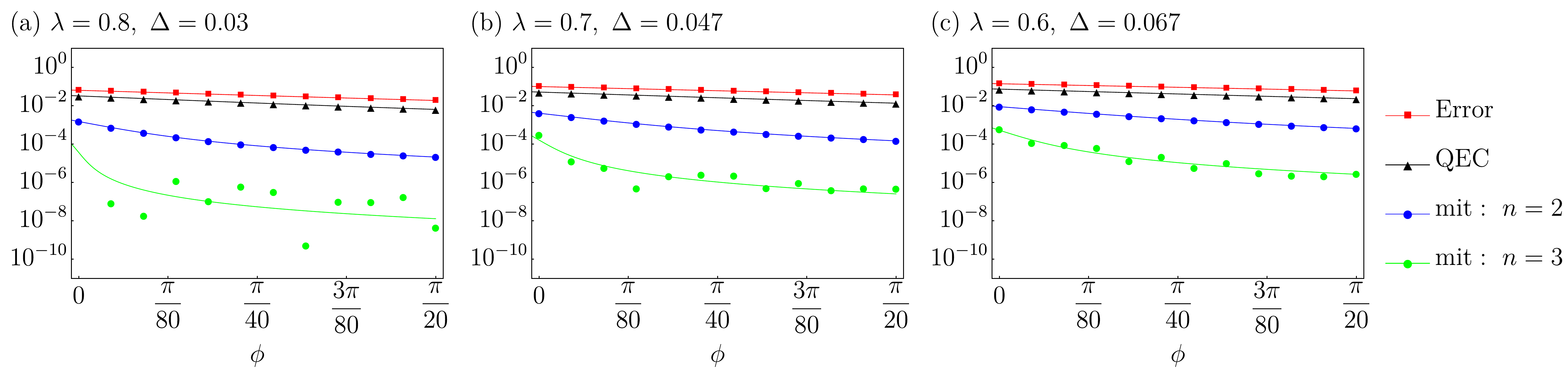}
    \caption{(a)-(c) 
    The square of biases (with a logarithmic scale) of the $\phi$ estimation, where $\phi \in \left[0, \frac{\pi}{20}\right]$, exploiting $5$-qubit Twin graph state with two pairs of true twins in the presence of the local depolarizing noise. Other features are consistent with those in Fig. \ref{fig:GHZDephasing}.
    } 
    \label{fig:TwinPauli}
\end{figure*}
Lastly, we note that for $5$-qubit twin graph case, even though the code distance of $\mathcal{D}(\mathbf{s}^{[\text{c}]})$ is not larger than $2$, we find that $B_{\text{mit}}$ can be reduced up to $O(\Delta^{3})$.

\subsection{$7$-qubit Steane code}

\begin{figure*}[t]
    \centering
    \includegraphics[width=\textwidth]{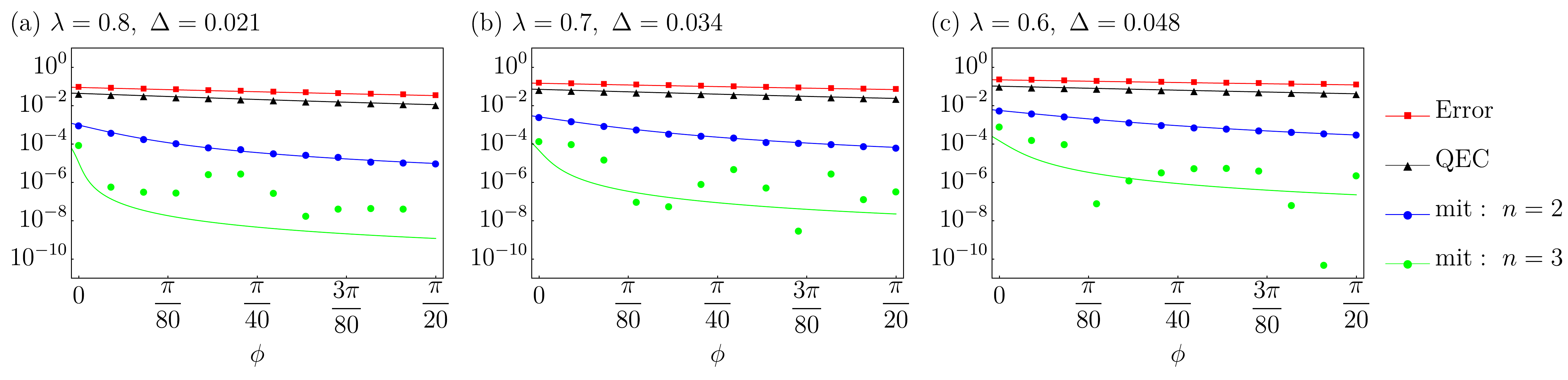}
    \caption{(a)-(c) 
     The square of biases (with a logarithmic scale) of the $\phi$ estimation, where $\phi \in \left[0, \frac{\pi}{20}\right]$, exploiting logical $0$ state in the $7$-qubit Steane code in the presence of the local depolarizing noise. Other features are consistent with those in Fig. \ref{fig:GHZDephasing}.
    } 
    \label{fig:SteanePauli}
\end{figure*}

In the last scenario (in Fig. \ref{fig:SteanePauli}), we consider the logical $0$ state of the $7$-qubit Steane code where the corresponding CGSs and NCGSs are \begin{align}
    &\mathbf{s}^{[\text{c}]}=\{\hat{Z}^{(2)}\hat{Z}^{(3)}\hat{Z}^{(4)}\hat{Z}^{(5)},\hat{Z}^{(4)}\hat{Z}^{(5)}\hat{Z}^{(6)}\hat{Z}^{(7)},\hat{Z}^{(1)}\hat{Z}^{(3)}\hat{Z}^{(5)}\hat{Z}^{(7)},\hat{Z}^{(1)}\hat{Z}^{(2)}\hat{Z}^{(3)}\hat{Z}^{(4)}\hat{Z}^{(5)}\hat{Z}^{(6)}\hat{Z}^{(7)}\},\\
    & \mathbf{s}^{[\text{n}]}=\{\hat{X}^{(2)}\hat{X}^{(3)}\hat{X}^{(4)}\hat{X}^{(5)},\hat{X}^{(4)}\hat{X}^{(5)}\hat{X}^{(6)}\hat{X}^{(7)},\hat{X}^{(1)}\hat{X}^{(3)}\hat{X}^{(5)}\hat{X}^{(7)}\}.
\end{align}
Here, $\hat{A} \equiv \hat{X}^{(4)}\hat{X}^{(5)}\hat{X}^{(6)}\hat{X}^{(7)}$ is chosen as the measurement observable. While this quantum probe does not achieve quantum-enhanced precision, we study this scenario for pedagogical purposes.

\begin{figure*}[t]
    \centering
    \includegraphics[width=\textwidth]{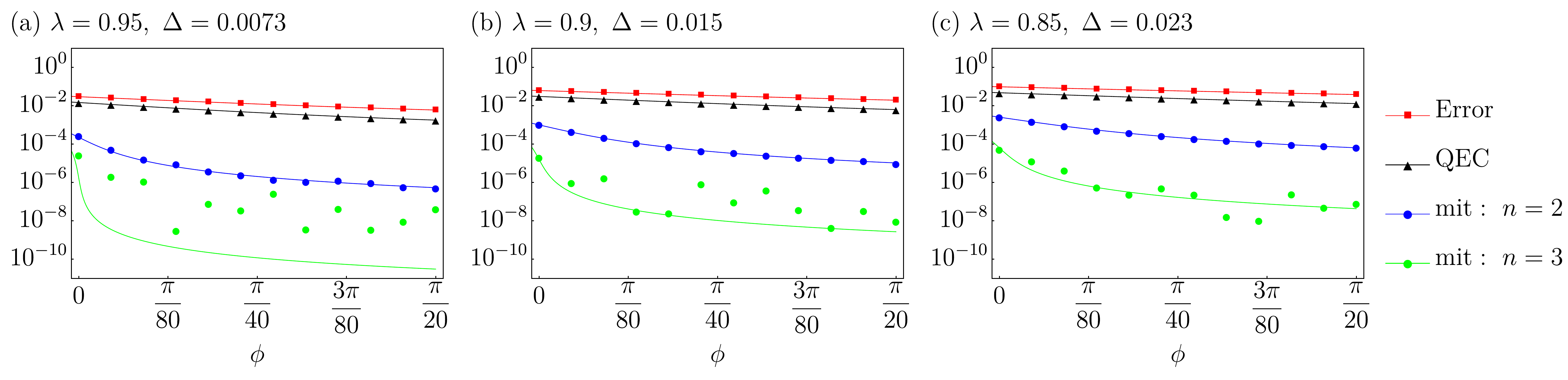}
    \caption{(a)-(c) 
     The square of biases (with a logarithmic scale) of the $\phi$ estimation, where $\phi \in \left[0, \frac{\pi}{20}\right]$, exploiting logical $0$ state in the $7$-qubit Steane code in the presence of the local dephasing noise. Other features are consistent with those in Fig. \ref{fig:GHZDephasing}.
    } 
    \label{fig:SteaneDephasing}
\end{figure*}

\setcounter{section}{0}

\end{document}